\documentclass[11pt]{article}
\pdfoutput=1
\usepackage{jheppub}
\usepackage{upgreek}
\usepackage{float, extarrows, tikz-cd}
\usepackage{graphicx}
\usepackage{subfig}
\usepackage{slashed}
\usepackage{tabularx,ragged2e}
\usepackage{longtable}
\usepackage{amssymb}
\usepackage{amsmath,amssymb}
\usepackage{slashed}
\usepackage{caption}
\usepackage{xcolor}
\usepackage{dsfont}
\usepackage{verbatim}
\usepackage{mathtools,xcolor,ytableau,amsfonts,tikz}
\usepackage{physics}
\usepackage{simpler-wick}
\usepackage{microtype}
\usepackage[nobottomtitles*]{titlesec}
\usepackage{color,soul}
\usepackage{cancel}
\setulcolor{orange}
\newcolumntype{C}{>{\Centering\arraybackslash}X}

\usepackage{array}
\newcolumntype{P}[1]{>{\centering\arraybackslash}p{#1}}

\numberwithin{equation}{section}

\let\oldfootnote\footnote
\renewcommand{\footnote}[1]{%
    \begingroup%
    \linespread{1}
    \oldfootnote{#1}%
    \endgroup%
}

\newcommand{\mL}{\mathcal{L}}
\newcommand{\mO}{\mathcal{O}}
\newcommand{\pd}{\partial}
\newcommand{\mR}{\mathcal{R}}
\newcommand{\Gb}{\overline{G}}

\newcommand\Dmin{\Delta_{\text{min}}}

\newcommand\beq{\begin{equation}}
\newcommand\eeq{\end{equation}}

\author[a,b]{Gabriel Cuomo,}
\author[c]{Leonardo Rastelli,}
\author[d]{Adar Sharon}

\affiliation[a]{Center for Cosmology and Particle Physics, Department of Physics, New York University, New York, NY 10003, USA}
\affiliation[b]{Department of Physics, Princeton University, Princeton, NJ  08544, USA}
\affiliation[c]{C. N. Yang Institute for Theoretical Physics, Stony Brook University, Stony Brook, NY 11794, USA}
\affiliation[d]{Simons Center for Geometry and Physics, SUNY, Stony Brook, NY 11794, USA}

\emailAdd{gc6696@princeton.edu}			 
\emailAdd{leonardo.rastelli@stonybrook.edu}
\emailAdd{asharon@scgp.stonybrook.edu}    

\title{Moduli Spaces in CFT: Large Charge Operators}

\abstract{Using the large-charge expansion, we prove a 
necessary condition for a CFT to exhibit conformal symmetry breaking,
under the assumption that a continuous global symmetry is {\it also} broken on the moduli space: there must be a tower of charged local operators
whose scaling dimensions are asymptotically linear in the charge.
In supersymmetric theories with a continuous R-symmetry and a holomorphic moduli space, the existence
of such a tower of operators follows trivially from a BPS condition: their scaling dimensions are then exactly linear in the R-charge. 
We illustrate the more general statement in several examples of three-dimensional ${\cal N}=1$ CFTs, where the leading linear behavior
receives nontrivial corrections. By considering a suitable scaling limit, we also relate the spectrum of  states with large charge on the cylinder
(isomorphic to local operators) 
to the spectrum of massive particles on the moduli space. 
}

\begin{document}

\maketitle

\section{Introduction}

Conformal field theories that exhibit spontaneous breaking of conformal symmetry (a moduli space of vacua) are highly  non-generic. Their key property, the absence of a potential for the dilaton, is a special feature that requires fine tuning. It seems however
difficult to fully characterize this feature 
in terms of abstract CFT data. In principle, one can set up an abstract bootstrap problem, involving the usual
data (scaling dimensions and OPE coefficients) as well as new data such as the spectrum of
asymptotic states in the broken vacuum and form factors. 
In a recent paper~\cite{Cuomo:2024vfk}, we illustrated
the simplest bootstrap equation \cite{Karananas:2017zrg}
in a concrete perturbative example.\footnote{See~\cite{Ivanovskiy:2024vel} for a similar analysis in ${\cal N} =4$
SYM theory.} Our findings confirmed the general intuition that theory data must satisfy some very special relations
for conformal symmetry breaking to occur, but
a simple general criterion is still elusive.

In this work we address the problem from a different angle.
We prove a simple {\it necessary}
condition for spontaneous conformal symmetry breaking to occur, under the (crucial) assumption that a continuous global symmetry
is {\it also} broken on the 
moduli space. 
Let $\Delta_{\rm min} (Q)$ be the minimum scaling dimension among all local operators
of charge\footnote{Here for ease of notation we take  the global symmetry to be $U(1)$.} $Q$. The necessary condition is that $\Delta_{\rm min} (Q)$ must be asymptotically linear,
\begin{equation} \label{criterion}
\text{CFT with moduli space s.t. }\cancel{Q}\;\implies\;
\Delta_{\rm min} (Q) = \alpha_0 Q + \dots \quad {\rm for}\, Q \to \infty \, ,
\end{equation}
where the dots indicate possible subleading corrections.

It will come as no big surprise that
our basic tool is the large-charge 
expansion initiated in~\cite{Hellerman:2015nra},\footnote{
This is by now a large subject, see~e.g.~\cite{Gaume:2020bmp} for a review.}
a beautiful and surprising  application of effective field theory (EFT) ideas. 
Even in a  strongly-coupled CFT, certain observables of large charge $Q\gg 1$ are captured by a weakly-coupled EFT, for which $1/Q$ is the natural small parameter. One such observable is precisely $\Delta_{\text{min}}(Q)$, which 
  by the state operator correspondence is
  given by the ground-state energy  on the cylinder of the large-charge EFT at fixed $Q$.
Our claim (\ref{criterion}) follows 
by a straightforward extension of the analysis of~\cite{Hellerman:2017veg}, enforcing the sole additional assumption  that the theory has a moduli space where the charge is
spontaneously broken. The form of the
 subleading corrections
 is dimension dependent. 
In $d=3$ we find the behavior
\begin{equation} \label{3dbehavior}
    \Delta_{\text{min}}(Q)=\alpha_0 Q +\alpha_1+\alpha_2/Q+O(1/Q^2)\;,
\end{equation}
while in $d=4$ the expansion takes the form 
\begin{equation}
    \Delta_{\text{min}}(Q)=\alpha_0 Q +\alpha_1\log Q+\alpha_2+O(1/Q)\;.
\end{equation}

The necessary condition
(\ref{criterion}) is established in full generality, with no recourse to supersymmetry.
However, when it comes to check it 
in concrete examples, we  must face the predicament that
  all {\it known}
 interacting, local\footnote{In most of the paper we assume the existence of a local stress tensor with a finite two-point function. It is not difficult to find examples
 of CFTs with an approximate moduli space at leading in a large $N$ expansion,
 as we briefly discuss in 
 section \ref{sec:largeN}.}
 CFTs with a moduli space of vacua are in fact supersymmetric.  The most familiar class of examples have at least four real supercharges in $d=3$ and $d=4$ spacetime dimensions. Their moduli spaces are holomorphic, parametrized by the vacuum expectation values of chiral operators. Chiral operators are charged under a 
continuous R-symmetry, which is thus also spontaneously broken. In these theories, holomorphy goes hand in hand with the existence of infinite towers of protected chiral operators with {\it exactly} linearly-spaced scaling dimensions, proportional to the R-charge, $\Delta (Q) = \alpha Q$. Mathematically, the ring of chiral operators coincides with the holomorphic 
coordinate ring on the moduli space.\footnote{The more precise statement is that the coordinate ring is identified with {\it reduced} chiral ring, i.e.~with the ring of chiral operators where nilpotents elements have been quotiented out. This can be established rigorously in Lagrangian models~\cite{Luty:1995sd} and it is believed to be true (general lore!)  in any 3d and 4d SCFT with at least four real supercharges.} Physically, the chiral operators
arise (at least at some intuitive level) from quantization of the holomorphic moduli. In this whole class of examples, (\ref{criterion}) follows trivially from a BPS condition. 
From the viewpoint of the large-charge EFT,
this amount of supersymmetry imposes selection rules that set to zero the higher coefficients $\alpha_{i\geq 1}$.

Fortunately, there is another class of supersymmetric  theories
for which (\ref{criterion}) is nontrivial: the
 three-dimensional SCFTs with only ${\cal N}=1$ SUSY (two real supercharges). These theories have no continuous R-symmetry, no holomorphy and no protected operators. Nevertheless, 
they can still admit exact moduli spaces thanks a  discrete $\mathbb{Z}_2$ R-symmetry~\cite{Gaiotto:2018yjh}, as we review in appendix \ref{app:epsilon_exp}. 
They are thus the perfect playground  to study the interplay between moduli spaces and global symmetries.

We consider several explicit examples of
$\mathcal{N}=1$ SUSY theories with moduli spaces, and check that 
 (\ref{criterion}) holds  with non-vanishing corrections to the leading linear behavior. 
 In selecting the matter content and interactions  of 
 these examples, some care had to be taken to ensure that
 they have a moduli space  with a broken global charge while at the same time  no  emergent ${\cal N}=2$ supersymmetry in the IR.
 Most of our examples are Wess-Zumino theories, which 
can be studied  perturbatively in  $4-\epsilon$ dimensions
by analytically continuing in the number of fermions~\cite{Fei:2016sgs, ThomasSeminar},
as we review in appendix~\ref{app:epsilon_exp}.
We go beyond the EFT analysis and compute explicitly the coefficients $\alpha_i$  in the $\epsilon$-expansion,  following \cite{Badel:2019oxl}.

In our concrete examples, the leading correction to the linear behavior of $\Delta_{\text{min}}(Q)$ turns out to be negative or zero. This is perhaps related to the charge convexity conjecture \cite{Aharony:2021mpc}, which schematically claims that $\Delta_{\text{min}}(Q)$ is a convex function of $Q$. While a counterexample to the original conjecture appeared in \cite{Sharon:2023drx}, our findings might support a modification of the conjecture 
(``convexity only at large enough charge'') that bypasses the counterexample. However we have not been able to prove this weaker conjecture and leave
further exploration  to future work. 

The large-charge EFT allows us  compute additional observables as well, including OPE coefficients and dimensions of operators just above $\Delta_{\text{min}}(Q)$. 
Further predictions can be obtained
 by assuming the existence of a macroscopic limit \cite{Jafferis:2017zna}. This limit relates correlation functions in the large-charge state on the cylinder to flat-space correlators on the moduli space by taking $Q\to\infty$ as well as taking the cylinder radius $R\to\infty$ by keeping a specific ratio fixed. Physically, this limit corresponds to ``zooming in'' to distances much smaller than the sphere radius but keeping the charge large, so that we expect the result to be a flat-space correlator in a nontrivial state. By applying this procedure to e.g.~four-point functions of large-charge operators, we can relate the spectrum of large charge operators to the spectrum of massive particles on the moduli space.

We can summarize the main conceptual lesson as follows.
In
CFTs with a moduli space where a global charge is also broken,
large charge operators create, in radial quantization, semiclassical states that closely resemble the  vacua where both the conformal and the internal symmetry are broken.
This implies the existence of a tower of operators whose scaling dimension grows linearly with the charge.
The existence of a conserved charge is essential, as it allows to focus on a precise tower of states, those corresponding to the lowest dimensional charged operators. However one might wonder whether 
more generally, in CFTs with a moduli space but {\it no} global charge,
there could  still be some simple semiclassical signature
of spontaneous conformal symmetry breaking  in the high energy spectrum.
In section \ref{sec:comments} we offer some musings on this question,
starting from the EFT of a real dilaton.
We speculate that in the general case the moduli space is reflected in the existence of certain special resonant states on the cylinder, whose width becomes parametrically narrower than their energy in the high energy limit.

\bigskip
\noindent
The remainder of the paper is organized as follows. In section \ref{sec:EFT_and_CFT} we review the construction of the moduli space EFT and we compute $\Delta_{\text{min}}(Q)$ as  well as of some additional observables. We explain in detail how to extract the leading-order result   and the form of the corrections to it. Next in section \ref{sec:examples} we discuss several explicit examples of moduli spaces at large charge in $d=3$ ${\cal N}=1$ theories.
We compute the first correction to the leading-order result 
and discuss some applications.
We also briefly consider examples of non-supersymmetric large $N$ theories with an approximate moduli space.
In section~\ref{sec:macroscopic} we apply the macroscopic limit to large-charge correlators and relate them to flat-space correlators on the moduli space. This limit allows us to write down precise equations relating the spectrum of large charge local operators to the massive spectrum on the moduli space. In section~\ref{sec:comments} we offer some speculations on the general case of moduli spaces
with no broken global symmetry. Several appendices complement the text with further  technical details.

\bigskip 

\bigskip

\section{EFT and CFT data at large charge}\label{sec:EFT_and_CFT}

In this section we  argue that if a CFT admits a moduli space of vacua in which a continuous internal symmetry group $G$ is (either partially or completely) spontaneously broken, then the lowest dimensional operator with large charge under a broken Cartan generator $Q$ has scaling dimension linearly proportional to the charge: 
\begin{equation}\label{eq_moduli_to_Delta}
\text{CFT with moduli space s.t. }\cancel{Q}\;\implies\;\Delta_{\rm min}(Q)\propto Q \quad {\rm for}\, Q \to \infty \, .
\end{equation}
For non-abelian groups, $\Delta_{\rm min}(Q)$ is the scaling dimension of an operator in a large representation of $G$.

The argument leading to this result uses the EFT for the massless modes on the moduli space and the state-operator correspondence. This approach is a simple generalization of that of \cite{Hellerman:2017veg}, where the authors studied a specific supersymmetric theory, and the lowest dimensional charged operators have protected dimensions $\Delta(Q) = \alpha Q$ due to a BPS shortening condition.  As it will become clear below,  the conclusion that $\Delta_{\rm min}(Q)\propto Q$ for large $Q$ is independent of the details of the theory and of  supersymmetry, as long as the CFT admits a moduli space. The proportionality coefficient in \eqref{eq_moduli_to_Delta}, as well as the subleading corrections and the spectrum of large-charge operators
with $\Delta \gtrsim \Delta_{\rm min}$
are also calculable from the EFT.

\subsection{Moduli effective theory}\label{sec:moduliEFT}

Consider a CFT with an $n$-dimensional moduli space in $d>2$. We denote its internal symmetry group $G$, and assume that along the moduli space the conformal symmetry is spontaneously broken along with the internal symmetry group as $G\to H$, for some (possibly trivial) $H\subset G$.
We can use an EFT to describe states on a moduli space at energies much smaller than the gap $m$ of the massive states. The massless degrees of freedom always include at least $n$ scalar fields whose expectation values locally parametrize the moduli space; among these, we always find a dilaton $\Phi$ together with $\text{dim}(G/H)$ Goldstone bosons. In supersymmetric theories the massless degrees of freedom also include the superpartners. 
At a generic point on the moduli space
it is always possible to make a change of coordinates such that there is one noncompact direction (which we associate with the dilaton $\Phi$) and $n-1$ compact directions $\phi_A$, $A=1,...,n-1$ corresponding to the additional moduli, including the Goldstones for the internal symmetry.

The EFT is constrained by the nonlinear realization of the conformal symmetry and the spontaneously broken internal symmetries \cite{Salam:1969bwb,Isham:1970gz,Ellis:1971sa}. In practice, the EFT is most easily constructed by imposing Weyl invariance and introducing a Weyl invariant combination of the metric and the dilaton as (see e.g. \cite{Luty:2012ww})
\begin{equation}\label{eq_g_hat}
\hat{g}_{\mu\nu}=g_{\mu\nu} |\Phi|^{\frac{4}{d-2}}\,,
\end{equation}
where we parametrized the dilaton $\Phi$ so that it transforms as a real canonically normalized scalar field in $d$-dimensions under the conformal group, and 
we take the additional moduli $\phi^A$ dimensionless, and thus with zero Weyl weight.
In this section $g_{\mu\nu}$ can be taken to be the flat metric $g_{\mu\nu}=\text{diag}(1,-1,-1,-1)$, but note that we could also work on any other Weyl equivalent manifold by Weyl invariance; we will make use of this observation in the next section. 

Denoting with a hat geometric invariants constructed from the rescaled metric~\eqref{eq_g_hat}, it is easy to write the effective action to leading order in derivatives 
\begin{multline}\label{eq_EFT}
S_{EFT}=\int d^dx\sqrt{\hat{g}}\left[\frac{1}{2}\hat{G}_{AB}(\phi)\hat{g}^{\mu\nu}\pd_\mu\phi^A\pd_\nu\phi^B +\hat{G}_{\Phi\Phi}(\phi)\frac{d-2}{8(d-1)} \hat{\mathcal{R}}
\right]\\+S_{\rm CFT_{\rm IR}}+\text{irrelevant couplings}\,,
\end{multline}
where $\hat{G}_{AB}(\phi)$ and $\hat{G}_{\Phi\Phi}(\phi)$ specify the metric of the non-linear sigma model (NLSM) and are constrained by the nonlinear realization of the internal symmetry or supersymmetry when present.  The coefficient multiplying $\hat{G}_{\Phi\Phi}$ was chosen for future convenience and the rescaled Ricci scalar contains the dilaton kinetic term
\begin{equation}\label{eq_dilaton_kinetic}
\hat{\mathcal{R}}=|\Phi|^{-\frac{4}{d-2}}\left[\mathcal{R}-\frac{4(d-1)\nabla^2\Phi}{(d-2)\Phi}\right]\,.
\end{equation}
Note indeed that in flat space the first line of the action~\eqref{eq_EFT} can be written as a standard~NLSM
\begin{equation}\label{eq_EFT_flat}
\int d^dx\left[\frac{1}{2}G_{AB}(\Phi,\phi)\pd_\mu\phi^A\pd^\mu\phi^B +\frac{1}{2}G_{\Phi\Phi}(\phi)(\pd\Phi)^2+G_{A\Phi}(\Phi,\phi)\pd_\mu\phi^A\pd^\mu\Phi \right] \,,
\end{equation}
where conformal invariance determines the dependence of the metric on $\Phi$ and relates $G_{\Phi\Phi}$ and $G_{A\Phi}$.
\begin{equation}\label{eq_EFT_NLSM_metric}
G_{AB}(\Phi,\phi)=\Phi^2\hat{G}_{AB}(\phi)\,,\quad
G_{\Phi\Phi}(\phi)=\hat{G}_{\Phi\Phi}(\phi)\,,\quad
G_{A\Phi}(\Phi,\phi)=\frac12\Phi\frac{\pd}{\pd \phi^A}\hat{G}_{\Phi\Phi}(\phi)\,.
\end{equation}
In~\eqref{eq_EFT} $S_{\rm CFT_{\rm IR}}$ schematically denotes all the other massless IR fields, including the superpartners and/or accidental strongly coupled sectors.  The Poincar\'e invariant vacua with broken conformal symmetry correspond to constant solutions of the form
\begin{equation}\label{eq_modulis_vev}
\langle \Phi\rangle=v^{\frac{d-2}{2}}\,,\qquad
\langle\phi^A\rangle=\text{const}\,.
\end{equation}
Expanding in fluctuations around such solutions we find that the moduli correspond to $n$ weakly coupled massless scalar particles. Note that by dimensional analysis the gap of the massive modes, and hence the cutoff of the EFT, scales as $m\propto v$.

All the interactions of the moduli among themselves and/or with the additional fields present in $S_{\rm CFT_{\rm IR}}$ are irrelevant.  From the expansion of the action~\eqref{eq_EFT_flat} we see that the moduli are derivatively coupled.  In the simplest case where no strongly coupled sector is present,  additional interactions arise from the couplings to the superpartners (if present) and from higher derivative terms.  The latter are suppressed by inverse powers of $|\Phi|^{\frac{2}{d-2}}\sim v$ by Weyl invariance. For instance, based solely on conformal invariance the first higher derivative operators arise at fourth order in derivatives and are given by
\begin{equation}
    \label{eq_EFT_4derivatives}
\begin{split}
S_4  =&\int d^dx\sqrt{\hat{g}}\left[c_1(\phi)(\hat{\mR}_{\mu\nu})^2+c_2(\phi)(\hat{\mR}_{\mu\nu\rho\sigma})^2+c_{3,AB}(\phi)\hat{\mR}^{\mu\nu}\pd_\mu\phi^A\pd_\nu\phi^B \right.
\\
 +& \, \left.c_{4,ABCD}(\phi)\hat{\mR}^{\mu\nu\rho\sigma}
\pd_\mu\phi^A\pd_\nu\phi^B\pd_\rho\phi^C\pd_\sigma\phi^D+
c_{5,ABCD}(\phi)(\pd_\mu\phi^A\hat{\nabla}^\mu\phi^B)(\pd_\nu\phi^C\hat{\nabla}^\nu\phi^D)\right]\,, 
\end{split}
\end{equation}
where we used the leading order equations of motion to eliminate operators proportional to $\hat{\mR}$ and $\hat{\nabla}^2\phi^A$. The components $\hat{G}_{AB}$ and $\hat{G}_{\Phi\Phi}$ of the NLSM metric in~\eqref{eq_EFT}, as well as the functions $c_i$'s appearing in~\eqref{eq_EFT_4derivatives} are Wilson coefficients, whose value depends on the microscopic dynamics of the specific underlying CFT. For an underlying strongly coupled theory, these scale as inverse powers of $(4\pi)$ according to generalized dimensional analysis~\cite{Georgi:1992dw} (at a generic point of the moduli space), while weakly coupled theories correspond instead to non-generic sizes for these coefficients (generically in the form of a large $\sim 1/g^2$ prefactor).

In even dimensions,  higher derivative operators include also Wess-Zumino terms required to match the conformal anomaly and  other 't Hooft anomalies of the internal symmetries.  In particular, in $d=4$ conformal anomaly matching implies that the EFT should include the following four-derivatives term \cite{Schwimmer:2010za,Komargodski:2011vj}:
\begin{equation}\label{eq_EFT_WZ}
S_{\rm WZ}=-\int d^dx\log\left(|\Phi|^{\frac{2}{d-2}}/\mu\right) \left(\Delta a \, E_4-\Delta c \, W^2_{\mu\nu\rho\sigma}\right)+\ldots\,,
\end{equation}
where $\mu$ is an arbitrary mass scale, $\Delta a$ and $\Delta c$ denote the difference between the UV and IR anomaly coefficients,  $E_4$ and $W_{\mu\nu\rho\sigma}$ are, respectively, the Euler density and the Weyl tensor, and the dots stand for terms with at least two derivatives acting on $\Phi$, which are required by the Wess-Zumino consistency conditions and whose specific form will not be important for our purposes. In general in $2n$-dimensions, the Wess-Zumino anomaly matching term is of order $2n$ in derivatives \cite{Baume:2013ika}.

It is important to remark that the absence of relevant couplings for the moduli is a very special property. Indeed, based solely on conformal invariance, the dilaton always admits at least a relevant coupling,  the cosmological constant term:
\begin{equation}\label{eq_cc_dilaton}
\int d^dx\sqrt{\hat{g}}f(\phi)=\int d^dx\sqrt{g}\, |\Phi|^df(\phi)\,,
\end{equation}
where the function $f$ is constrained by the internal symmetry and may just be a constant.
This term, as well as other potential couplings to relevant operators of $S_{\rm CFT_{\rm IR}}$, would create a potential for $\Phi$ which would lift the flat direction. Therefore, for the CFT to admit a moduli space, its selection rules must set all the relevant couplings of the dilaton including~\eqref{eq_cc_dilaton} to zero.  This is a highly nontrivial condition, and typically requires some amount of supersymmetry.\footnote{An alternative scenario, considered in \cite{Coradeschi:2013gda}, would be to have a family of theories labeled by some marginal couplings, such that a flat direction opens up at special points of the conformal manifold.} As elaborated in~\cite{Coradeschi:2013gda} (see also~\cite{Bellazzini:2013fga}), the problem of constructing interacting non-supersymmetric CFTs with a moduli space of vacua can be thought as a scalar analogue of the cosmological constant problem.  In our analysis, we will simply use self-consistently the general EFT~\eqref{eq_EFT}, remaining agnostic about the mechanism that produces the moduli space and tunes the cosmological constant term to zero.

The EFT breaks down when the metric of the sigma model $\sim v^{d-2}\hat{G}(\bar\phi)$ becomes singular. This obviously happens at $v=0$,  but it may also happen at other isolated points or surfaces of dimension $p<n$, at which additional degrees of freedom become light. At those points we may still use an EFT of the form~\eqref{eq_EFT} for the $n-p$ moduli \emph{parallel} to the surface, formally including the contribution of the additional light sector in $S_{\rm CFT_{\rm IR}}$. A familiar illustration of this behaviour is in $\mathcal{N}=4$ SYM with gauge group $SU(N)$. At a generic point of the moduli space the gauge group is fully Higgsed and the low energy theory consists of $N$ $\mathcal{N}=4$ $U(1)$ gauge theories, but at specific points the gauge group is partially restored and the massless modes include nontrivial sectors with gauge group $SU(p<N)$.\footnote{
A less trivial example is a $3d$ $\mathcal{N}=1$ theory of $11$ real superfields with superpotential
\begin{equation}\label{eq_W_11_fields}
W=g_1A X_iY_i+g_2B X_i Z_i+g_2C Y_iU_i\,,
\end{equation}
where $i=1,2$. This model admits a fixed-point in the $\epsilon$-expansion, and its symmetries protect a moduli space which
includes an $O(2)$ breaking branch
\begin{equation}\label{eq_11_field_moduli}
\langle X_i\rangle= \hat{n}_i v_x \,,\quad \langle Y_i\rangle=\varepsilon_{ij} \hat{n}_j v_y\,,\quad
\langle U_i\rangle= \hat{n}_i v_u \,,\quad
\langle Z_i\rangle=\varepsilon_{ij} \hat{n}_j v_z\,,
\end{equation}
where $\hat{n}_i$ is a unit vector. 
The moduli space~\eqref{eq_11_field_moduli} is isomorphic to $\mathds{R}^4\times S^1$ and the low energy EFT at a generic point consists of $5$ derivatively coupled real superfields. This manifold is singular when $v_x=0$ or $v_y=0$, where the low energy EFT includes 8 moduli 
as well as an interacting sector. 
Within this surface we encounter an additional singularity when three of the vevs in~\eqref{eq_11_field_moduli} are zero, where there are nine light modes, an axio-dilaton and an interacting 
SCFT. \label{footnote_11_fields}
}

\subsection{Derivation of the leading order result}

Let us suppose that in some region of the moduli space the internal symmetry $G$ is spontaneously broken to $H$. Denoting with $Q_a$ the broken Cartan charges, 
without loss of generality we can parametrize the coset $G/H$ 
such that the action of $Q_a$ simply amounts to a shift of the associated Goldstone field $\pi^a$.  This implies that $\hat{G}_{AB}(\phi)$ and $\hat{G}_{\Phi\Phi}(\phi)$ in~\eqref{eq_EFT} can only depend on the $\pi^a$'s through derivative terms. From now on we shall distinguish the $\pi^a$'s from the other moduli, that will be denoted $\phi^A$ as before with a slight abuse of notation.

In the following it will be convenient to work on the Lorentzian cylinder $\mathds{R}\times S^{d-1}$, for which the moduli action reads
\begin{equation}\label{eq_EFT_cyl}
\begin{split}
S_{EFT}=\int d^dx\sqrt{g}&\left[\frac{\Phi^2}{2}\pd_\mu\pi^a\pd^\mu\pi^b \hat{G}_{ab}(\phi)+\Phi^2\pd_\mu\pi^a\pd^\mu\phi^B \hat{G}_{aB}(\phi)\right.
\\
&\left.
+\frac{\Phi^2}{2}\pd_\mu\phi^A\pd^\mu\phi^B \hat{G}_{AB}(\phi)
+\frac{\hat{G}_{\Phi\Phi}(\phi)}{2}\left(-\Phi\pd^2\Phi-m_d^2\Phi^2 \right)
\right]+\ldots\,,
\end{split}
\end{equation}
where we separated the $\pi^a$'s explicitly from the other moduli and  
\begin{equation}
m_d^2=\frac{d-2}{4(d-1)}\mathcal{R}=\frac{(d-2)^2}{4 R^2} \,.   
\end{equation}
is the conformal mass coupling on the cylinder and $R$ denotes the radius of $S^{d-1}$.

While the EFT is derived around the background~\eqref{eq_modulis_vev}, the low energy effective action may be reliably used around arbitrary backgrounds as long as the derivative expansion is under control. We shall therefore search for a solution of the equations of motion on the cylinder with the ansatz
\begin{equation}\label{eq_cyl_sol}
\Phi=v^{\frac{d-2}{2}}\,,\qquad
\pi^a=m^a t/R\,,\qquad
\phi^A=\bar{\phi}^A\,,
\end{equation}
where $t$ denotes the time coordinate.  The ansatz~\eqref{eq_cyl_sol} ensures that the expectation values of the current and the stress-tensor agree with the structure of a three-point function of primary operators \cite{Cuomo:2021ygt}; therefore~\eqref{eq_cyl_sol} describes a charged primary state.\footnote{This can also be checked pedestrianly by computing the expression for the special conformal generators in terms of the fluctuations around the background.} Physically, the ansatz~\eqref{eq_cyl_sol} describes a stationary solution where both time translations $D/R$ (corresponding to dilations in flat space) and the charges $Q_a$ are spontaneously broken,\footnote{Of course, strictly speaking there is no spontaneous symmetry breaking at finite volume and integrating over the zero-modes of the solution~\eqref{eq_cyl_sol} the symmetry is restored at a quantum level (see e.g.  the discussion in \cite{Monin:2016jmo}).} leaving invariant a diagonal combination $D-m_aQ_a$: this is the spontaneous symmetry breaking pattern defining a superfluid phase \cite{Son:2002zn,Nicolis:2015sra}, that recently attracted much attention in the study of large charge operators in CFTs, see e.g. \cite{Hellerman:2015nra,Monin:2016jmo,Gaume:2020bmp}. In generic CFTs however the \emph{chemical potentials} $ m^a$ grow with the charge and the scaling dimension scales as $\Delta_{\rm min}(Q)\propto Q^{\frac{d}{d-1}}$. The existence of the moduli space (or in other words the absence of a potential for the dilaton~\eqref{eq_cc_dilaton}) makes the models at hand non-generic, and we will find below that the $m^a$'s do not scale with $Q$, resulting in a linear relation between the charge and the scaling dimension.

The equation of motion (EOM) for $\Phi$ imposes the following condition (for $v\neq 0$)
\begin{align}\label{eq_PHI_eom}
m^a m^b \hat{G}_{ab}(\bar{\phi})/R^2-\hat{G}_{\Phi\Phi}(\bar{\phi})m_d^2 =0\,,
\end{align}
while the $\bar{\phi}_A$ satisfy the following condition
\begin{equation}\label{eq_phi_EOM}
\left[m_d^2\frac{\pd \hat{G}_{\Phi\Phi}(\phi)}{\pd \phi^A}-\frac{m^a m^b}{R^2}\frac{\pd \hat{G}_{ab}(\phi)}{\pd\phi^A}\right]_{\phi=\bar{\phi}}=0\,.
\end{equation}
Finally, computing the Noether currents $j^\mu_a$, we find the values of the Cartan charges on the solution~\eqref{eq_cyl_sol}
\begin{equation}\label{eq_Q_fix}
j^0_a=\frac{Q_a}{R^{d-1}\Omega_{d-1}}=\hat{G}_{ab}(\bar{\phi})m^b v^{d-2}/R\,,
\end{equation}
where $\Omega_{d-1}=2\pi^{d/2}/\Gamma(d/2)$ is the volume of the $d-1$-dimensional sphere. Equations~\eqref{eq_PHI_eom},~\eqref{eq_phi_EOM} and~\eqref{eq_Q_fix} completely specify the solution~\eqref{eq_cyl_sol} in terms of the charges $Q_a$.

Let us consider first the case in which only one charge $Q_a=\delta_a^1 Q$ is non-zero (in some basis).  We can thus solve~\eqref{eq_Q_fix} for $v$ as
\begin{equation}\label{eq_v_Q}
v=\frac{1}{R}\left(\frac{Q}{\Omega_{d-1} \hat{G}_{1b}m^b}\right)^{\frac{1}{d-2}}\,.
\end{equation}
The remaining equations do not depend on $v$ nor $Q$.  For fixed values of the $\bar{\phi}^A$ such that $\hat{G}_{AB}$ and $\hat{G}_{\Phi\Phi}$ are non-singular,  it is always possible to find $m^a$'s that solve \eqref{eq_PHI_eom} and~\eqref{eq_Q_fix} as a function of the $O(1)$ ratio $\hat{G}_{AB}/\hat{G}_{\Phi\Phi}$.\footnote{The explicit solution for arbitrary charges reads
\begin{align}
 v^{2d-4}=\frac{j_a^0(\hat{G}^{-1})^{ab}j_b^0}{\hat{G}_{\Phi\Phi}m^2_d}\,,
\qquad
\frac{m^a}{R}= \frac{(\hat{G}^{-1})^{ab}j_b^0}{v^{d-2}}=
\frac{ (\hat{G}^{-1})^{ab}j_b^0}{\sqrt{j_a^0(\hat{G}^{-1})^{ab}j_b^0/(\hat{G}_{\Phi\Phi}m^2_d)}}\,.
\end{align}
} Our only assumption is that~\eqref{eq_phi_EOM} also admits a solution at a point $\bar{\phi}^A$ that lies within EFT - we shall comment further about this hypothesis later.   The energy of the state is obtained from the energy momentum tensor: 
\begin{equation}\label{eq_Deltamin_1}
\Delta_{\rm min}(Q)=R\int d^{d-1}x\sqrt{g}\,T_{00}=R\int d^{d-1}x\sqrt{g}\left(j^0_a m_a/R-\mathcal{L}\right)=m_1 Q\,,
\end{equation}
where we used that~\eqref{eq_PHI_eom} implies that the Lagrangian vanishes, $\mL=0$, and $m_1\sim O(1)$ depends upon the Wilson coefficients of the EFT through the equations of motion.  By the state-operator correspondence, \eqref{eq_Deltamin_1} is the scaling dimension of the lowest dimensional charged operator of the theory.  When the solution to~\eqref{eq_phi_EOM} is not unique, the minimal energy one has to be taken (i.e. the one that gives the smallest $m_1$) and in some cases there might be several degenerate operators with the same charge and scaling dimension at this order. 

The generalization of the argument to operators in different representations, whose highest weight state has multiple non-zero Cartan charges, is trivial and yields
\begin{equation}\label{eq_Deltamin_many}
\Delta_{\rm min}(\{Q_a\})=m_a Q_a=Q f\left(Q_a/Q_b\right)\,,\qquad
Q=\left(\sum_a Q_a^2\right)^{1/2}\,,
\end{equation}
where the function $f$ is again determined by the EFT Wilson coefficients. Similarly to~\eqref{eq_Deltamin_1}, \eqref{eq_Deltamin_many} holds as long as $Q\gg 1$. The function $f$ is expectedly analytic except at isolated points in the space span by the ratios $Q_a/Q_b$, corresponding to singularities in the moduli space.

Our analysis is reliable as long as the dilaton vev is much larger than the derivatives of the Goldstone fields $v\sim Q^{\frac{1}{d-2}}/R\gg |\pd\pi_a|\sim m_a/R$; we conclude that the result~\eqref{eq_Deltamin_1} holds for $Q\gg 1$.  At the classical level subleading contributions arise from higher derivative terms and are suppressed by powers of $R^2v^2\sim Q^{\frac{2}{d-2}}$. For instance, it is simple to check that the operators in~\eqref{eq_EFT_4derivatives} yield a contribution of order $Q^{\frac{d-4}{d-2}}$ to~\eqref{eq_Deltamin_1}.  Note also that the Wess-Zumino term~\eqref{eq_EFT_WZ} evaluates to zero on the cylinder and therefore does not contribute to the scaling dimension.  

For a purely weakly-coupled EFT $Q$ is the loop counting parameter and thus quantum corrections to the result~\eqref{eq_Deltamin_1} (generically) scale as $Q^0$. These are more important than classical higher derivative contributions for $d\leq 4$. We will discuss such contributions and the spectrum of operators corresponding to excitations on top of the large charge ground state in the next subsection.

It is perhaps useful at this point to discuss a trivial example: the free theory of a charged scalar $\varphi$. In this case the moduli space EFT coincides with the full action:
\begin{equation}
\mL=|\pd\varphi|^2-m_d^2|\varphi|^2\,.
\end{equation}
The parallel with~\eqref{eq_EFT_cyl} becomes obvious in cylindrical coordinates $\varphi=\Phi e^{-i\pi}/\sqrt{2}$. In this case the solution~\eqref{eq_cyl_sol} is
\begin{equation}\label{eq_free_th_sol}
\varphi=e^{-i m_d t}v^{\frac{d-2}{2}}\,.
\end{equation}
The charge and energy densities yield the scaling dimension
\begin{equation}\label{eq_free_th_Delta}
j^0=2 m_d v^{d-2}\,,\quad
T_{00}= m_d j^0\quad\implies\quad
\Delta_{\rm min}(Q)=\frac{d-2}{2 }Q\,,
\end{equation} 
which is obviously exact in this case.

Besides free theories, there are many other instances for which the relation~\eqref{eq_Deltamin_1} is exact. This is generally the case for theories with BPS operators, where $Q$ is one of the $R$-charges. In these cases supersymmetry ensures the absence of subleading corrections even if the EFT is generically nontrivial. In some case, e.g. for Coulomb branch operators in theories with rank larger than one, there are many degenerate solutions, corresponding to different directions in the chiral ring. We will discuss examples in which the relation~\eqref{eq_Deltamin_1} is not related to a BPS shortening condition in the next section.

Note that, while the cylinder makes the physical properties of the solution~\eqref{eq_cyl_sol} manifest, it is also possible to work directly in flat space or any other conformally flat manifold by Weyl invariance; this is for instance illustrated by the semiclassical saddle-point calculation of the flat space two-point function $\langle\varphi^n\bar{\varphi}^n\rangle$ in free theory, presented, e.g., in~\cite{Badel:2019oxl}.

Let us finally come back to the solution of~\eqref{eq_phi_EOM}. Suppose that no solution for the $\bar{\phi}^A$'s exist at nonsingular points of the moduli space. In this case we can still consider the configuration~\eqref{eq_cyl_sol} for some $\bar{\phi}^A$ such that $G_{AB}$ and $c$ are non-singular. We therefore find a non-minimal energy state with $\Delta(Q)=m_1 Q$ for some $m_1$. We conclude that the real ground state, even if it does not lie within EFT, must have lower energy: $\Delta_{\rm min}(Q)<m_1 Q$.

In practice we expect that,  when the solution to~\eqref{eq_phi_EOM} lies at a singular submanifold of the moduli space, we can simply use the low energy theory along that surface for the $n-p$ parallel moduli to compute the scaling dimension. Such EFT is formulated in terms of different Wilson coefficients than those at a generic point of the moduli space, but we nonetheless recover the linearity of $\Delta_{\rm min}(Q)$ at large charge.

\subsection{Nearby operators}

We now use the EFT to compute the spectrum of excited states on top of the large charge ground states as well as quantum corrections on top of the result~\eqref{eq_Deltamin_1}.

Let us begin analyzing the spectrum of excited states. To compute the scaling dimension of operators associated with excitations of the moduli we need to compute the action for the fluctuations around the solution~\eqref{eq_cyl_sol}. We do this in appendix~\ref{app:fluctuations}.  We find that the fluctuations always include two modes with dispersion relation
\begin{equation}\label{eq_EFT_2modes}
\omega_{B,1}(\ell)=\ell/R\,,\qquad
\omega_{B,2}(R)=(\ell+d-2)/R\,,
\end{equation}
as a function of the angular momentum $\ell$.  Note that~\eqref{eq_EFT_2modes} coincides with the spectrum of fluctuations around the solution~\eqref{eq_free_th_sol} in free theory.  The dispersion relations for the additional $n-2$ fields are also approximately linear in the limit $\ell\gg 1$:
\begin{equation}\label{eq_EFT_n2_modes_large_l}
\omega_{B,k}(\ell)=\ell/R+O(1/R)\,,\qquad k=3,\ldots,n\,.
\end{equation}
The general form of the dispersion relations~\eqref{eq_EFT_n2_modes_large_l} at finite $\ell$ may be quite involved due to the spontaneous breaking of time translations in~\eqref{eq_cyl_sol}; we were not able to obtain general closed form expressions. A simplification occurs when there is a unique Cartan charge and a symmetry acting as $\pi\rightarrow-\pi$ on the corresponding Goldstone boson.  This is for instance the case in theories with a $U(1)$ charge invariant under charge conjugation. In this case we find a fully relativistic result
\begin{equation}\label{eq_EFT_n2_modes_relativistic}
\omega_{B,k}(\ell)=\sqrt{n_k^2/R^2+J_{\ell}^2}\,,
\end{equation}
where $J^2_{\ell}=\ell(\ell+d-2)/R^2$ and the $n_k$'s are $O(1)$ numbers that depend upon the coefficients in~\eqref{eq_app_fluct_action} of the appendix.

The Fock space of these quasi-particles describes operators in the same representation of the internal symmetry with higher scaling dimension, except for the zero modes of $\omega_{B,1}$ and the other $n_{G/H}-1$ Goldstone bosons, that relate different charge sectors or create states in the same representation of the ground state. States with one or more spin $1$ $\omega_{B,1}$ particles, whose gap is $\omega_{B,1}(1)=1/R$, are descendants.

At a generic point of the moduli space, in most cases, there are additional massless modes. In the simplest and most generic scenario, these consist of derivatively coupled massless fermions and vector fields (with Abelian gauge group in $d\leq 4$), these often being the superpartners of the moduli.\footnote{Note that, unless the solution~\eqref{eq_cyl_sol} describes a BPS state, the dispersion relation of these modes is not directly fixed by the supersymmetry.} The excitations of these fields describe additional operators.  Below, we briefly analyze these additional states.

Assuming four-component Dirac fermions in $2<d\leq 4$ for concreteness (the generalization to chiral spinors and other dimensions is straightforward), the most general fermionic action compatible with conformal invariance at first order in derivatives reads
\begin{equation}\label{eq:Sfermions}
S_{\rm fermions}\simeq\int d^dx\sqrt{\hat{g}}\left\{\frac{i}{2}\bar{\psi}^i G_{\psi,ij}(\phi)\left[\hat{\slashed{\nabla}}\psi^j+\Gamma^j_{ka}(\phi)\gamma^\mu\psi^k\pd_\mu\pi^a
+\Gamma^j_{kA}(\phi)\gamma^\mu\psi^k\pd_\mu\phi^A\right]+c.c.\right\}\,,
\end{equation}
where the hatted metric is defined in~\eqref{eq_g_hat}, the indices $i,j$ run from $1$ to the number of fermions $N_F$ and the \emph{metric} and \emph{connections} in field space $G_{\psi}$ and $\Gamma$ are model-dependent. 
Note that terms of the form $\sqrt{\hat{g}}\bar{\psi}\psi$ would yield a large mass $\sim v\bar{\psi}\psi$ in the broken vacua~\eqref{eq_modulis_vev}, as well as a potentially relevant Yukawa coupling. We therefore assumed that such operators are ruled out by the selection rules of the model similarly to~\eqref{eq_cc_dilaton}.  Around the solution~\eqref{eq_cyl_sol} the quadratic action reads
\begin{equation}\label{eq_EFT_fermions}
S_{\rm fermions}\simeq v\int d^dx\sqrt{g}\left\{\frac{i}{2}\bar{\psi}^i G_{\psi,ij}(\bar{\phi})\left[\slashed{\nabla}\psi^j+m^a\Gamma^j_{ka}(\bar{\phi})\gamma^0\psi^k \right]+c.c.\right\}\,.
\end{equation}
We see that the coupling to the Goldstones $\pi^a$ creates a chemical potential. Therefore, canonically normalizing the fields and decomposing them in spinor harmonics (see e.g.~\cite{Camporesi:1995fb}), we find that the dispersion relations take the form
\begin{equation}\label{eq_EFT_fermions_disp_rel}
\omega_{F,i}^{(\pm)}(\ell)=p_F(\ell)\pm \tilde{\mu}_i\,,
\end{equation}
where $p_F(\ell)=\left[\ell+(d-1)/2\right]/R$ and the $\tilde{\mu}_i$ are $O(1/R)$ chemical potentials that depend on the Wilson coefficients $G_{\psi}$ and $\Gamma_{\psi}$ in~\eqref{eq_EFT_fermions}.  
Each mode in~\eqref{eq_EFT_fermions} has spin $\ell+1/2$ and must be counted twice for a four component Dirac field (besides the group-theoretical multiplicity). Interestingly, \eqref{eq_EFT_fermions_disp_rel} implies that when one or more of the $\tilde{\mu}_i$'s becomes larger than some of the $ p_{F}(\ell)$ we have negative energy levels. The fermions must therefore occupy such states, forming a (small) Fermi sphere that contributes an $O(1)$ amount to the charges $Q_a$. The gap of the Fock states around the Fermi sphere is given by the absolute value of~\eqref{eq_EFT_fermions_disp_rel}: $|\omega_{F,i}(\ell)|$.

In practical examples the fermionic metric and connection $G_\psi,\Gamma$ in \eqref{eq:Sfermions} are related to the bosonic NLSM~\eqref{eq_EFT_cyl} by supersymmetry. A relevant example for us is $3d$ $\mathcal{N}=1$ SUSY theories.
In this case the generic EFT consists of $n$ real superfields, with bosonic action as in~\eqref{eq_EFT} and fermionic action given by:\footnote{The second term in~\eqref{eq_EFT_fermion_3dSUSY} can also be written in terms of the connection associated with the NLSM~\eqref{eq_EFT_NLSM_metric} (using that the $\psi$'s are Grassmanian).}
\begin{equation}\label{eq_EFT_fermion_3dSUSY}
   S_{\rm fermions}= -\frac{i}{2}\int d^3x\,\psi^{i}\sigma^\mu\left[G_{ij}(\varphi)\pd_\mu +\partial_{\mu}\varphi^{k}\frac{\partial G_{kj}(\phi)}{\partial\varphi^i}\right]\psi^{j}+\ldots
\end{equation}
where the $\psi^i$'s are Majorana, $i=1,\ldots, n$, $\varphi^i=\{\Phi,\pi^a,\phi^A\}$ collectively denotes all the moduli, the NLSM metric is the one given in~\eqref{eq_EFT_NLSM_metric} and the ellipsis denote 4-fermion interactions. The bosonic 2-derivative terms thus completely determine the fermionic two-derivative terms through SUSY.

Finally, the two derivative action for the vector fields at leading order is  simply the free theory action accounting for the conformal coupling to the dilaton\footnote{An apparent exception in $d=3$ is the following coupling $F_{\mu\nu}\varepsilon^{\mu\nu\rho}\pd_{\rho}\phi^A f_{V,A}(\phi)$ for an Abelian vector field, but this is not a new term compared to~\eqref{eq_EFT_flat} since an Abelian vector is dual to a shift invariant scalar.}
\begin{equation}
S_{\rm vectors}\simeq -\frac{1}{4}\int d^dx\sqrt{\hat{g}}\,G_{V,ij}(\phi)\hat{g}^{\mu\rho}\hat{g}^{\nu\sigma}F_{\mu\nu}^iF_{\rho\sigma}^j\,,
\end{equation}
where the field strength is defined as usual and the function $G_{V,ij}(\phi)$ is constrained by gauge-invariance for non-Abelian gauge groups (that can occur in $d>4$). Expanding around the solution~\eqref{eq_cyl_sol} we find that each vector field behaves as a free Maxwell field and thus we find $N_V$ modes with dispersion relation \cite{Giombi:2015haa}:
\begin{equation}
\omega_{V,i}(\ell)=\sqrt{(\ell+1)(\ell+d-3)}/R\,,
\end{equation}
where $\ell\geq 1$.

\subsection{Quantum corrections}

Let us now discuss the quantum corrections to the result~\eqref{eq_Deltamin_1}. In the generic case where all the fields in the EFT are derivatvely coupled, these arise from the Casimir energy due the quasi-particles analyzed above:
\begin{equation}\label{eq_EFT_Casimir}
\frac{R}{2}\left[\sum_{k=1}^n \sum_{\ell=0}^\infty n_B(\ell)\omega_{B,k}(\ell)-2\sum_{k=1}^{N_F}\sum_{\ell=0}^{\infty}\sum_{\pm} n_F(\ell)|\omega_{F,k}^{(\pm)}(\ell)|+
\sum_{k=1}^{N_V} \sum_{\ell=1}^\infty n_V(\ell)\omega_{V,k}(\ell)\right]\,,
\end{equation}
where the multiplicities are given by
\begin{equation}\label{eq_multiplicities}
\begin{gathered}
n_B(\ell)=\frac{(d+2 \ell-2) \Gamma(d+\ell-2)}{\Gamma(d-1) \Gamma(\ell+1)}, \quad n_F(\ell)=\frac{2 \Gamma(d+\ell-1)}{\Gamma(d-1) \Gamma(\ell+1)} \,,\\[0.5em]
n_V(\ell)=\frac{\ell (\ell+d-2) (2 \ell+d-2) \Gamma (\ell+d-3)}{\Gamma (d-2) \Gamma (\ell+2)}\,.
\end{gathered}
\end{equation}
The sum in~\eqref{eq_EFT_Casimir} of course must be properly regularized compatibly with the symmetries of the model. \eqref{eq_EFT_Casimir} naively provides a $Q^0$ contribution to the scaling dimension.  For odd spacetime dimensions, the result must be finite since it cannot be renormalized by any counterterm. For even $d$ however we find higher derivative operators that contribute to the scaling dimension at order $Q^0$ compatibly with conformal invariance, such as the operators in~\eqref{eq_EFT_4derivatives} for $d=4$. Therefore in even dimension we expect that, barring special symmetries of the solution,  the sum~\eqref{eq_EFT_Casimir} displays a logarithmic divergence and thus results in a logarithm of the cutoff $v\sim Q^{\frac{1}{d-2}}$ similarly to the discussion in \cite{Cuomo:2020rgt}. In summary:
\begin{equation}\label{eq_EFT_Casimir_result}
\delta\Delta^{(\rm 1-loop)}_{\rm min}(Q)=\begin{cases}
\alpha_0 & d\;\text{odd} \\
\alpha_0\log Q & d\;\text{even}\,,
\end{cases}
\end{equation}
where $\alpha_0$ is a constant that depends upon the dispersion relations. \eqref{eq_EFT_Casimir_result} is the leading correction to~\eqref{eq_Deltamin_1} in $d=3,4$. When multiple Cartan charges are non-zero as in~\eqref{eq_Deltamin_many}, the coefficient $\alpha_0$ in~\eqref{eq_EFT_Casimir_result} is a function of the ratios $Q_a/Q_b$.

Of course, in many cases supersymmetry enforces cancellations. E.g. for BPS states we always have $\delta\Delta^{(\rm 1-loop)}_{\rm min}(Q)=0$. We will discuss in the next sections examples for which the Casimir energy is nontrivial.

In generic schemes,  such as dimensional regularization\footnote{Note that in the presence of a dilaton there exist many perturbative schemes that explicitly preserve conformal symmetry~\cite{Gretsch:2013ooa}, including a version of dimensional regularization.}, \eqref{eq_EFT_Casimir} can be further simplified.  Indeed, the first two scalar modes in~\eqref{eq_EFT_2modes} do not contribute to the sum (as it can be easily checked in dimensional regularization) because their dispersion relations coincide with those for the fluctuations of a free scalar field. For the same reason, the contribution of the fermions in~\eqref{eq_EFT_fermions_disp_rel} vanishes unless one or more of the $\tilde{\mu}_i$ are larger than some of the $p_F(\ell)$, in which case the fermions' contribution equals the (negative) energy of the Fermi sphere.\footnote{Note that the Casimir energy~\eqref{eq_EFT_Casimir} is the first correction to $\Delta_{\rm min}(Q_a)-m^aQ_a$, where the $Q_a$'s include the $O(1)$ contributions from the Fermi surface.} Finally the vector fields contribution to~\eqref{eq_EFT_Casimir} is trivial in $d=4$ since the cylinder is a conformally flat manifold and the four dimensional Maxwell action is Weyl invariant on its own, without the need of a dilaton.

Finally, we estimate the subleading corrections to $\Delta_{\rm min}(Q)$ when $S_{\rm CFT_{\rm IR}}$ in~\eqref{eq_EFT_flat} includes some arbitrary interacting sector.\footnote{This discussion might apply in the example discussed in footnote~\ref{footnote_11_fields} if the saddle describing the lowest dimensional charged operator lies in one of the singular submanifolds discussed below~\eqref{eq_11_field_moduli}.} Recall that by assumption this sector couples to the moduli only via irrelevant terms.  This means that when expanding around~\eqref{eq_modulis_vev} all operators are irrelevant. As we show below, this does not ensure that around a different background, such as~\eqref{eq_cyl_sol}, no relevant perturbation arises. We will nonetheless argue that such terms only yield subleading contribution to~\eqref{eq_Deltamin_1}, therefore proving the robustness of our result.

If $S_{\rm CFT_{\rm IR}}$ contains a primary scalar operator $\mO_{\Delta}$ of dimension $\Delta<d-2$, excluding total derivatives the leading couplings around the background~\eqref{eq_cyl_sol} are given by
\begin{equation}\label{eq_Odelta}
S\supset\int d^dx\sqrt{\hat{g}}\left[\gamma_{ab}(\phi)\pd_\mu\pi^a\hat{\pd}^\mu\pi^b
+\beta(\phi)\hat{\mR}\right]\mO_{\Delta}|\Phi|^{-2\frac{\Delta}{d-2}}\,,
\end{equation}
where the powers of the dilaton are dictated as usual by conformal invariance. Recalling~\eqref{eq_g_hat} and using~\eqref{eq_cyl_sol}, \eqref{eq_Odelta} results in a perturbation of the form
\begin{equation}\label{eq_Odelta2}
\sim M^{d-\Delta}\int d^dx\sqrt{g}\mO_{\Delta}\,,
\end{equation}
where by~\eqref{eq_v_Q} the mass scale upfront is given by
\begin{equation}\label{eq_M_Odelta}
M= \left[\frac{v^{d-2-\Delta}}{R^2}\right]^{\frac{1}{d-\Delta}}\sim\frac{1}{R}Q^{\frac{d-2-\Delta}{(d-\Delta)(d-2)}}\,.
\end{equation}
As announced, \eqref{eq_Odelta2} is a relevant perturbation for $\Delta<d$, and the coupling~\eqref{eq_M_Odelta} is larger than the geometric scale $1/R$ for $\Delta<d-2$. Therefore the coupling~\eqref{eq_M_Odelta} triggers a nontrivial RG flow. This results (at most) in a contribution $\sim M^d$ to the potential. Using~\eqref{eq_M_Odelta} we conclude that~\eqref{eq_Odelta} implies the existence of the following contribution to the scaling dimension:
\begin{equation}\label{eq_dDelta_strong}
\delta\Delta_{\rm min}(Q)\sim Q^{1-\frac{2\Delta}{(d-\Delta)(d-2)}}\,,
\end{equation}
which is larger than the Casmir energy~\eqref{eq_EFT_Casimir_result} for $\Delta<d-2$, but is smaller than the leading term in~\eqref{eq_Deltamin_1} for $\Delta>0$.\footnote{Of course $\Delta>0$ is always realized in unitary theories, but we must have $\Delta>0$ also in non-unitary theories for the term~\eqref{eq_Odelta2} to be irrelevant in the Poincar\'e' invariant state~\eqref{eq_modulis_vev}.} In other words, the contribution of the coupling~\eqref{eq_Odelta} to the dilaton potential is always subleading compared to the conformal coupling to the Ricci, as expected.\footnote{As a concrete illustration of this mechanism we can imagine that $S_{\rm CFT_{\rm IR}}$ consists of a four-dimensional scalar $\varphi$ with potential $\lambda\varphi^4$ (this coupling can be made conformal compensating the dependence on the RG scale with the dilaton \cite{Komargodski:2011xv}). Identifying therefore $\mO_{\Delta}=\varphi$ in~\eqref{eq_Odelta}, with the appropriate sign the scalar acquires a vev $\langle\varphi\rangle\sim M$ and at its minimum the potential contributes to the energy as $\sim Q^{2/3}$, up to logarithms of $Q$ associated with the running of $\lambda$, in agreement with~\eqref{eq_dDelta_strong}.}

When $S_{\rm CFT_{\rm IR}}$ does not include scalars of dimension $\Delta<d-2$, assuming unitarity the only non-irrelevant coupling between the interacting sector and the moduli can take the form
\begin{equation}
S\supset\int d^dx\sqrt{\hat{g}} \lambda_a(\phi)\pd_\mu\pi^a J^\mu|\Phi|^{-\frac{2d}{d-2}}\simeq \int d^dx\sqrt{g} \lambda_a(\bar{\phi})m^a J^0/R\,,
\end{equation}
where $J^\mu$ is a conserved current of $S_{\rm CFT_{\rm IR}}$. As for the chemical potential term in the fermionic action~\eqref{eq_EFT_fermions}, this operator may result in a $\sim Q^0$ contribution to $\Delta_{\rm min}(Q)$. Similarly, a marginally irrelevant operator would couple to the dilaton through the coupling beta-function
\cite{Komargodski:2011xv}, and might result in a contribution to the scaling dimension proportional to some inverse power of $\log Q$.

In conclusion, accidental strongly coupled sector might yield corrections to the scaling dimension $\Delta_{\rm min}(Q)$ which are non-polynomial in $Q$, but are nonetheless suppressed with respect to the leading order result~\eqref{eq_Deltamin_1}.

\subsection{Correlation functions}\label{sec:correlation_functions}

Let us conclude the general discussion by mentioning that the EFT can also be used to compute correlation functions with two or more large charge operators and an arbitrary number of light operators as in the non supersymmetric case~\cite{Monin:2016jmo}. For instance, we can represent a neutral operator of scaling dimension $\Delta \sim O(1)$ in the EFT in terms of the dilaton as
\begin{equation}\label{eq_EFT_op_matching}
\mO_{\Delta}=c_{\mO}(\phi)\Phi^{\frac{2\Delta}{d-2}}+\ldots\,,
\end{equation}
where $c_{\mO}$ is some function of the moduli and the dots stand for higher derivative terms. Evaluating its expectation value in the large charge ground state we obtain the OPE coefficient
\begin{equation}\label{eq_EFT_OPE}
\lambda_{Q\mO Q}=\langle Q|\mO_{\Delta}|Q\rangle\simeq c_{\mO}(\bar{\phi})v^{\Delta}\quad\implies\quad \lambda_{Q\mO Q}\propto Q^{\frac{\Delta}{d-2}}\,.
\end{equation}
Similar results hold for charged operators. For spinning operators one has to introduce derivatives of the Goldstones, and thus multiply the formula~\eqref{eq_EFT_op_matching} by factors of $\pd_\mu\pi^a/\Phi^{\frac{2}{d-2}}$. One obtains
\begin{equation}\label{eq_EFT_OPE_spin}
\lambda_{Q\mO Q}\vert_{\text{spin }J}\propto Q^{\frac{\Delta -J}{d-2}}\,,
\end{equation}
barring additional selection rules. Note that~\eqref{eq_EFT_OPE_spin} is obviously satisfied by the conserved currents. One may also consider correlation functions with several insertions of the large charge ground-state operator, as in \cite{Cuomo:2021ygt}.

Note that the result~\eqref{eq_EFT_OPE} holds in free theory, for which the OPE for $\langle\bar{\varphi}^{Q+q}\varphi^q\varphi^Q\rangle$ is $\sqrt{(Q+q)!}/\sqrt{Q!q!}\sim Q^{q/2}$ and $\Delta_{\phi^q}=q\frac{d-2}{2}$. Less trivially, the same results holds in strongly interacting $4d$ $\mathcal{N}=2$ SCFTs for Coulomb branch operators, as it follows from localization \cite{Grassi:2019txd,Beccaria:2020azj}, and for Higgs branch operators, whose correlators can be computed using the chiral algebra \cite{Beem:2013sza,Lemos:2020pqv}. The EFT calculation can also be extended to obtain subleading orders, see~\cite{Hellerman:2017sur,Hellerman:2018xpi,Hellerman:2020sqj,Hellerman:2021duh,Hellerman:2021yqz}. One can also check that~\eqref{eq_EFT_OPE_spin} agrees with the known chiral algebra result for the OPE between two Higgs branch operators and the spin $1$ Sch\"ur operator that is obtained fusing the $SU(2)_R$ current $J^{J_1J_2}_\mu$ and a Higgs branch operator $\mathcal{H}^{I_1\ldots I_q}$ in the $2q+1$ $SU(2)_R$ representation with $q\sim O(1)$.\footnote{This OPE is fixed because, within the chiral algebra,  the \emph{product} operator $\sim J^{J_1J_2}_\mu \mathcal{H}^{I_1\ldots I_q}$ is a Virasoro descendant of $\mathcal{H}^{I_1\ldots I_q}$.}

\section{Moduli spaces at large charge: examples}\label{sec:examples}

In this section we
test our basic claim (\ref{criterion}) in several nontrivial examples. To this end we focus 
on $3d$ $\mathcal{N}=1$ SCFTs, which have no continuous R-symmetry nor protected chiral operators (see appendix~\ref{app:epsilon_exp} for a brief review).
We will see explicitly how 
 the asymptotic linear behavior of $\Delta_{\rm min}(Q)$
receives subleading corrections.

We first review  how to compute the large charge scaling dimension $\Dmin(Q)$ in the presence of a small coupling $g$,
in a double-scaling limit where the mass of the single-particle states on the moduli space is held fixed.
We then study 
two specific Wess-Zumino models, where we can achieve perturbative control by working in $d=4-\epsilon$ dimensions,
such that $g^2\sim \epsilon$.
We are of course interested in models with a moduli space and at least a global $U(1)$ symmetry. The concrete Wess-Zumino examples that 
we consider may look somewhat baroque, but
some care needs to be taken in engineering models
that do
not exhibit an accidental supersymmetry enhancement
in the IR, with the global $U(1)$ becoming the continuous R-symmetry of an ${\cal N}=2$ SCFT.

We then consider the example of ${\cal N}=1$ SQED, going beyond perturbation theory using the effective action formalism discussed above. Finally we discuss some non-supersymmetric examples that exhibit approximate moduli spaces to leading order in a large $N$ expansion.

\subsection{Large charge double-scaling limit}

We quickly review the method for extracting large-charge data using the double-scaling limit discussed in \cite{Badel:2019oxl} (see also \cite{Watanabe:2019pdh}). Assuming a weak (cubic) coupling $g$ and a $U(1)$ symmetry, the method allows for the computation of $\Dmin(Q)$ in the double-scaling limit $g^2\rightarrow 0$ with  $g^2 Q=\text{fixed}$ and yields a result of the form
\begin{equation}\label{eq:double_scaling}
    \Dmin(Q)=\frac{1}{g^2} \hat{\Delta}_{-1}\left(g^2 Q\right)+\hat{\Delta}_0\left(g^2 Q\right)+g^2 \hat{\Delta}_1\left(g^2 Q\right)+\ldots\;,
\end{equation}
where the function $\hat{\Delta}_{-1}$, $\hat{\Delta}_0$, $\dots$ are, respectively, the tree-level, one-loop, and higher loop contributions.
When multiple charges are present, the functions $\hat{\Delta}_i$ also depend on the ratios $Q_a/Q_b$. In order to match the standard perturbative results, for small $g^2 Q$ the $\hat{\Delta}_i(g^2Q)$ admit a Taylor expansion in $g^2 Q$ which starts at order $g^2 Q$.  In all the examples we work in the $\epsilon$-expansion, such that $g^2\sim\epsilon$.

We use the state-operator correspondence, and compute the ground state energy at fixed charge $Q$ on the cylinder $\mathbb{R}\times S^{d-1}$. At small $g$ this can be done using a saddle-point approximation. These saddles take the form of the ``helical'' superfluid solution similar to~\eqref{eq_free_th_sol}, schematically
\begin{equation}\label{eq:saddle}
    X=\rho e^{i\mu t}
\end{equation}
where $X$ is a complex field on which the charge acts by phase rotations and $t$ is time. The constants $\rho$ and $\mu$ are fixed by the equations of motion and the charge-fixing condition. Then the leading-order term $\hat{\Delta}_{-1}$ is just the value of the energy computed on this saddle. 

The next-order contribution $\hat{\Delta}_0$ comes from the Casimir energy evaluated around the saddle. We must therefore quantize the theory around the saddle \eqref{eq:saddle} and compute the contribtions from the zero-point energies of the various modes. The regularization of the final infinite sum is straightforward within dimensional regularization, see \cite{Badel:2019oxl} for details. In principle one can use this formalism to compute higher orders in $\Dmin(Q)$ as well, although we will be content with focusing on these two leading contributions.

\subsection{7-field WZ model}\label{sec:7_field}

We study the model defined by the superpotential \cite{Gaiotto:2018yjh}
\begin{equation}\label{eq_uRu_def}
W=\frac{g}{2}\overline{u}\sigma^i u\, R_i\equiv \frac{g}{2}\overline{u}R u \;,
\end{equation}
where $u$ is a complex superfield which is a doublet under $SU(2)$ and $R$ is a real superfield and a triplet under $SU(2)$. The global symmetry group is $SU(2)\times U(1)$ where the $U(1)$ rotates the phase of $u$, and there is an additional $\mathbb{Z}_2^R$ symmetry taking $R\to -R$. The classical moduli space is generated by $R$ and is protected from quantum corrections. The model is IR dual to SQED with two charge-one particles \cite{Gaiotto:2018yjh,Benini:2018bhk}. 

In the $\epsilon$-expansion one finds the two-loop beta function 
\begin{equation}
    \beta_g=-\frac{g  \epsilon }{2}+\frac{3 g^3}{8 \pi ^2}-\frac{g^5}{8 \pi ^4}\;,
\end{equation}
leading to the two-loop anomalous dimensions\footnote{This was computed in \cite{Benini:2018bhk} apart from the $\epsilon^2$ term in $\Delta_{R^2}$, which is new.}
\begin{equation}
    \begin{split}
        \Delta_R&=1-\frac{\epsilon}{3}-\frac{\epsilon^2}{108}\;,\\
        \Delta_{R^2}&=2-\frac{2}{3}\epsilon-\frac{\epsilon^2}{54}\;,
    \end{split}
\end{equation}
where by $R^2$ we mean the symmetric product of two $R$'s. Surprisingly we see that $\Delta_{R^2}=2\Delta_R$ to two-loop order, which seems to hint at higher SUSY. However, this model is not compatible with emergent $\mathcal{N}=2$ SUSY, since the moduli space is 3-real-dimensional and so cannot be K{\"a}hler. Therefore the relation $\Delta_{R^2}=2\Delta_R$ cannot be exact and we should expect higher order to violate it. Below we will use the double-scaling limit analysis to analyze large charge operators.  We will also provide a structural explanation for why the relation $\Delta_{R^Q}=Q\Delta_R$ holds at two-loop, but not at higher orders.

To perform the analysis in the double-scaling limit with $g^2Q$ fixed, we first rewrite the superpotential in a more convenient form. Writing the fields in components as 
\begin{equation}
    u=\begin{pmatrix}  u_1 \\ u_2\end{pmatrix}\;,\qquad  R=\begin{pmatrix}R_3 & R_1+iR_2\\ R_1-iR_2 & -R_3
 \end{pmatrix}\;,
 \end{equation} 
the calculation consists of expanding around the saddle point $R_1+iR_2=ve^{-im_d t}$. To compute the leading terms $\hat\Delta_{-1},\,\hat\Delta_{0}$ in the expansion \eqref{eq:double_scaling}, it is enough to consider quadratic fluctuations around the saddle point (since $\hat\Delta_{-1}$ can be read off from the classical charges and $\hat\Delta_{0}$ is the Casimir energy around this background). The superpotential takes the form 
\begin{equation}\label{eq_uRu_pot_exp}
    W= \frac{g}{2}(R_1+iR_2)u_1^* u_2+c.c.+\frac{g}{2} R_3 (|u_1|^2-|u_2|^2)\;,
\end{equation}
and so in computing $\hat\Delta_{-1}$ and $\hat\Delta_{0}$ we can ignore the last term which is cubic around the saddle point. However, removing this term, the superpotential reduces to that of the $\mathcal{N}=2$ supersymmetric $XYZ$ model, which consists of three chiral superfields and superpotential 
\begin{equation}
    W=\frac{g}{2}XYZ\;.
\end{equation}
The theory thus exhibits enhanced SUSY at this order, which leads to some operator dimensions being protected. The $XYZ$ model was studied in the double-scaling limit in \cite{Sharon:2020mjs}, and the results are
\begin{equation}
    \frac{\hat{\Delta}_{-1}}{g_*^2}=\frac{d-2}{2} Q, \quad \hat{\Delta}_0=\frac{g_*^2 v^2}{8}=\frac{g_*^2 Q}{32 \pi^2}=\frac{\varepsilon}{6} Q\;.
\end{equation}
Summing up we find 
\begin{equation}\label{eq_uRu_Delta}
    \frac{\hat{\Delta}_{-1}}{g_*^2}+\hat\Delta_0=\frac{d-1}{3} Q\;,
\end{equation}
as expected from the BPS bound in the $XYZ$ model.

Equation~\eqref{eq_uRu_Delta} implies that $\Dmin(Q)=Q\Dmin(1)$ up to two-loop order in standard perturbation theory. This is because, as formerly mentioned, the functions $\hat{\Delta}_k(g^2 Q)$ in~\eqref{eq:double_scaling} are polynomials starting at linear order, and the deviation from $\Dmin(Q)=Q\Dmin(1)$ arises from the term $g^2 \hat\Delta_1$, which admits a Taylor expansion of the form $g^2\hat\Delta_1=\# g^4 Q+\#g^6Q^2+\ldots$ for small $g^2 Q$.

Finally notice that the mass of the gapped mode $u$ is proportional to $g^2 v^2\sim g^2 Q/R$. Therefore for $g^2 Q\gg 1$ we can integrate it out and obtain the moduli EFT.  We will analyze in detail such EFT in section~\ref{sec:SQED}.

In summary, we found that the large charge scaling dimension in the model~\eqref{eq_uRu_def} is linear in $Q$ as expected. Techinically, to the order we are working this result arises as a trivial consequence of the similarity with the $\mathcal{N}=2$ model studied in \cite{Sharon:2020mjs}, as we explained below~\eqref{eq_uRu_pot_exp} using the double-scaling limit. We will provide a less trivial check of our predictions in the next section.

\subsection{5-field WZ model}\label{subsec_5_field}

We study the WZ model defined by the superpotential
\begin{equation}\label{eq_5field_W}
    W=\frac{g}{2} A(|X|^2-|Y|^2)\;,
\end{equation}
with $A$ a real superfield and $X,Y$ complex superfields. Its continuous symmetries are $U(1)\times U(1)$, and in addition it has a discerete symmetry that flips the sign of $A$ and interchanges either $X$ and $Y$, or $X$ and $\bar{Y}$. Finally, the $\mathbb{Z}_2^R$ symmetry acts as $A\to -A$. The symmetries protect the classical moduli space (where either $\langle A\rangle\neq 0$ or $\langle |X|^2 \rangle=\langle |Y|^2 \rangle\neq 0$) from quantum corrections. According to our discussion, we expect operators with large charge under either of the $U(1)$'s to be well described by the effective action on the region of the moduli space where $X\neq 0$ and/or $Y\neq 0$,  where the internal symmetry group is either partially or fully broken. Below we perform a detailed study of the charged operators' spectrum of this theory within the $\epsilon$-expansion. This will provide a nontrivial check of the general predictions in section~\ref{sec:EFT_and_CFT}, concretely illustrating the relation between the moduli EFT and the large charge operators' spectrum.

In the $\epsilon$-expansion from $4d$ one finds the two-loop beta function
\begin{equation}
    \beta_g=-\frac{ \epsilon }{2}g+\frac{3 g^3}{32 \pi ^2}-\frac{3 g^5}{512 \pi ^4}\,,
\end{equation}
and the scaling dimensions
\begin{equation}\label{eq:pertU1XYdim}
\begin{aligned}
    \Delta_X=\Delta_Y&=1-\frac{5 \epsilon }{12}+\frac{ \epsilon ^2}{144}\,,\qquad & \Delta_A &=1-\frac{\epsilon }{3}+\frac{\epsilon ^2}{36}\;,\\
    \Delta_{X^2}=\Delta_{Y^2}&=2-\frac{2 \epsilon }{3}-\frac{\epsilon ^2}{36}\;,\qquad 
   & \Delta_{XY}=    \Delta_{X\bar{Y}}&= 2-\epsilon +\frac{\epsilon ^2}{36}\;,
    \\
    \Delta_{A X}=\Delta_{A Y}&=2-\frac{5 \epsilon }{12}+\frac{ \epsilon ^2}{144}\;,
\qquad
&\Delta_{|X|^2-|Y|^2}&=2-\frac{ \epsilon }{3}+\frac{\epsilon ^2}{36}\;,    
    \\
    \Delta_{A^2-|X|^2-|Y|^2}&=2-\epsilon+\frac{\epsilon ^2}{24} \;,\qquad
      &\Delta_{A^2+|X|^2+|Y|^2}&=
      2-\frac{\epsilon}{3}-\frac{\epsilon ^2}{8}\;.
\end{aligned}
\end{equation}
Using that to one-loop order the scaling dimension of the operator $X^{Q_x}Y^{Q_y}$ is a quadratic polynomial in $Q_x$ and $Q_y$, from~\eqref{eq:pertU1XYdim} we infer that the dimension of the lowest dimensional operator for arbitrary values of the charges is
\begin{equation}\label{eq_5field_Delta_pert}
\Delta_{X^{Q_x}Y^{Q_y}}=\Delta_{X^{Q_x}\bar{Y}^{Q_y}}=\left(Q_x+Q_y\right)\left(1-\frac{\epsilon}{2}\right)+\frac{\epsilon}{12}\left(Q_x-Q_y\right)^2+O\left(\epsilon^2\right)\,.
\end{equation}
This equation shows that to to one-loop order $\Delta_{X^{Q}Y^{Q}}=Q\Delta_{X Y}$. We will find below that this relation is exact at leading order in the large charge double-scaling limit, but it receives corrections at higher orders.

We now study the theory in the large-charge double-scaling limit. We start by computing the leading term $\hat\Delta_{-1}$ in the expansion \eqref{eq:double_scaling} for general charges $Q_x,\,Q_y$ (corresponding to the operators $X^{Q_x}Y^{Q_y}$). Without loss of generality we assume $Q_x\geq Q_y\geq 0$. The leading contribution is given by the semiclassical energy of the saddle of the form
\begin{equation}\label{eq_5field_saddle}
    X=\frac{v_x}{\sqrt{2}}e^{-i\mu_x t}\;,\quad Y=\frac{v_y}{\sqrt{2}} e^{-i\mu_y t}\;,
\end{equation}
with $A=0$. The EOMs and charge-fixing conditions fix
\begin{equation}\label{eq:EOMs}
\begin{aligned}
    \mu_x&=\sqrt{m_d^2+g^2 v_x^2-g^2 v_y^2}\;,\qquad
    & \frac{Q_x}{R^{d-1} \Omega_{d-1}}&= |v_x|^2 \mu_x\;,\\
   \mu_y&=\sqrt{m_d^2+g^2 v_y^2-g^2 v_x^2}\;, &
 \frac{Q_y}{R^{d-1} \Omega_{d-1}}&=|v_y|^2 \mu_y \;.
\end{aligned}
\end{equation}
\eqref{eq:EOMs} admit a unique solution for real values of $\mu_{x/y}$ and $|v_{x/y}|$. Interestingly, we find that the second derivative of the scaling dimension is discontinuous at $g^2Q_x/(R^{d-1}\Omega_{d-1})=8\sqrt{2}m_d$ and $Q_y=0$; the result it's otherwise analytic for $Q_x,\,Q_y>0$. The scaling dimension admits simple expansions for small and large values of $g^2 \sqrt{Q_x^2+Q_y^2}$
\begin{equation}\label{eq_5field_Delta_minus1}
    \frac{\hat{\Delta}_{-1}}{g^2 R}=\begin{cases}\displaystyle
m_{d}(Q_x+Q_y)+\frac{g^2(Q_x-Q_y)^2}{32R^{d-1}\Omega_{d-1}m_d^2}+\ldots & g^2\sqrt{Q_x^2+Q_y^2}\ll (4\pi)^2\\[1em]
\displaystyle
\sqrt{2}m_{d}\sqrt{Q_x^2+Q_y^2}\left[1
-\frac{\sqrt{2} m_d^4R^{d-1}\Omega_{d-1}(Q_x^2-Q_y^2)^2}{g^2(Q_x^2+Q_y^2)^{5/2}}
+\ldots 
\right]
& g^2\sqrt{Q_x^2+Q_y^2}\gg (4\pi)^2,
    \end{cases}
\end{equation}
which agrees with~\eqref{eq_5field_Delta_pert} for small $g^2 Q_{x/y}$. We were not able to obtain a closed form expression for $\hat{\Delta}_{-1}$ for arbitrary values of the charges, but we found that the result simplifies drastically for $Q_x=Q_y$, for which one surprisingly find that
\begin{equation}\label{eq_5field_Delta_minus1_equalQ}
    \frac{\hat{\Delta}_{-1}}{g^2}\stackrel{Q_x=Q_y=Q}{=}2 Rm_dQ=(d-2) Q\,,
\end{equation}
coinciding with the free theory result.

We now turn to the one-loop correction. We focus on the operators with $Q_x=Q_y=Q$ for which the leading order result~\eqref{eq_5field_Delta_minus1} is particularly simple. These states are special since they are invariant under the discrete symmetry exchanging $X$ and $Y$, and the solution~\eqref{eq_5field_saddle} explicitly reads
\begin{equation}\label{eq_5field_saddle_equalQ}
    \mu_x=\mu_y=m_d\,,\qquad
    |v_x|^2=|v_y|^2=\frac{Q}{m_d R^{d-1}\Omega_{d-1}}\,.
\end{equation}
Expanding the action to quadratic order around the saddle point, we find the dispersion relations given in appendix~\ref{app_5fields}.
Since the leading order result is independent of the coupling, we can neglect the renormalization of the coupling and the one-loop correction is just the Casimir energy
\begin{equation}\label{eq_5field_Casimir}
    \hat\Delta_0=\frac{R}{2} \sum_{k=1}^5 \sum_{\ell=0}^{\infty} \left[n_B(\ell) \omega_{B,k}(\ell)-2N_f \sum_{ \pm} n_F(\ell) \omega^{( \pm)}_{F,k}(\ell)\right]\;,
\end{equation}
where $ \omega_{B,k}$ and $ \omega_{F,k}$ are the bosonic and fermionic dispersion relations respectively, $N_f=1/4$ is the number of Dirac Fermions and the multiplicities are given in~\eqref{eq_multiplicities}. We checked that the sum~\eqref{eq_5field_Casimir} is regular (i.e. it contains no $1/\epsilon$ pole) in the limit $\epsilon\rightarrow 0$ when evaluated within dimensional regularization.

First, as a crosscheck, we compute the sum for small $g^2 Q$. We find that the Casimir energy admits an expansion starting at order $\epsilon^2 Q^2$
\begin{equation}\label{eq_5field_Delta0_small}
    \hat{\Delta}_0=\frac{g^4 Q^2}{4(4\pi)^4}+O\left(\frac{g^6 Q^3}{(4\pi)^6}\right)=
    \frac{\epsilon^2 Q^2 }{36}+O\left(\epsilon^3 Q^3\right)\,.
\end{equation}
The absence of an $O\left(\epsilon Q\right)$ correction is in agreement with the fact that both the double-scaling limit leading order result~\eqref{eq_5field_Delta_minus1_equalQ} and the one-loop diagrammatic calculation~\eqref{eq_5field_Delta_pert} coincide with the free theory answer.\footnote{Incidentally, comparing~\eqref{eq_5field_Delta0_small} with the diagrammatic result~\eqref{eq:pertU1XYdim} and using that the two-loop anomalous dimension of $X^QY^Q$ is a cubic polynomial with no $Q^0$ term, we obtain that the two-loop anomalous dimension (at fixed $Q$) is
\begin{equation}
    \Delta_{X^QY^Q}=\left(2-\epsilon\right)Q+\frac{\epsilon^2 Q^2 }{36}+O\left(\epsilon^3\right)\,.
\end{equation}}

For large $(g^2 Q)\gg (4\pi)^2$ we can evaluate the sum using the method of matched asymptotic expansion, as explained in appendix D of~\cite{Cuomo:2021cnb}. The leading terms of the result are
\begin{equation}
    \hat{\Delta}_0=\frac{g^2 Q}{64 \pi ^2}-\frac{1}{32} \log \left(\frac{g^2 Q}{4 \pi ^2}\right)+O\left(g^0Q^0\right)\,,
\end{equation}
and thus, summing the leading order contribution~\eqref{eq_5field_Delta_minus1_equalQ}, we arrive at
\begin{equation}\label{eq_5field_Delta_large}
    \frac{\hat{\Delta}_{-1}}{g^2}+\hat{\Delta}_0=\left[2-\frac{11 \epsilon }{12}+O\left(\epsilon^2\right)\right]Q-
    \frac{1}{32}\left[1+O\left(\epsilon\right)\right]\log\left(\epsilon Q\right)+O\left(\epsilon^0Q^0\right)
    ,.
\end{equation} 
\eqref{eq_5field_Delta_large} agrees with the general prediction of section~\ref{sec:EFT_and_CFT} that the scaling dimension grows linearly with the charge. This is a very nontrivial result at the technical level, and it is due to nontrivial cancellations between the bosonic and fermionic contributions (whose individual sums are divergent and scale as $\sim (g^2 Q)^2$ for large $g^2 Q$) in~\eqref{eq_5field_Casimir}. Note also that the coefficient of the linear term in~\eqref{eq_5field_Delta_large} deviates from the the free-theory result at this order.

Interestingly, the fist subleading contribution in~\eqref{eq_5field_Delta_large} at large $Q$ is logarithmic. This also agrees with the EFT prediction~\eqref{eq_EFT_Casimir_result}, since to leading order in $\epsilon$ the theory lives in four dimensions.  The coefficient of the logarithmic term is also predicted by the EFT. It is instructive to check this explicitly. 

For large charge $g^2\sqrt{Q_x^2+Q_y^2}\gg 1$ the fluctuations around the solution~\eqref{eq_5field_saddle} include two massive modes with gap $M^2\sim g^2(|v_x|^2+|v_y|^2)\sim g^2 \sqrt{Q_x^2+Q_y^2}/R^2\gg 1/R^2$. The moduli space is $\mathds{R}\times S^1\times S^1$, and therefore the low energy dynamics is well described by the EFT for the dilaton $\Phi$ and two $U(1)$ Goldstones $\pi_x$ and $\pi_y$ (and the superpartners). Using that the discrete symmetries act on the Goldstones as $\pi_x \leftrightarrow\pm\pi_y$, the most general bosonic effective action to leading order in derivatives reads
\begin{equation}
\mL_{bos}=\frac{c}{2}\Phi^2\left[(\pd\pi_x)^2 +(\pd\pi_y)^2 \right]+\frac12\left[(\pd\Phi)^2-m_d^2\Phi^2\right] \,,
\end{equation}
which depends on a unique Wilson coefficient $c$ to this order (we work in conventions such that the Goldstones are $2\pi$-periodic). The saddle-point describing an operator with charges $(Q_x,Q_y)$ is
\begin{equation}\label{eq_5field_EFT_saddle}
    \Phi=v\,,\quad\pi_x=\mu_x t\,,\quad\pi_y=\mu_y t\,,
\end{equation}
where the EOMs and the charge-fixing condition give
\begin{equation}
\mu_{x/y}=\frac{Q_{x/y}m_d}{\sqrt{c(Q_x^2+Q_y^2)}}\,,\qquad
v^2=\frac{\sqrt{Q_x^2+Q_y^2}}{\sqrt{c}\,m_d R^{d-1}\Omega_{d-1}}\,.
\end{equation}
Using~\eqref{eq_Deltamin_many}, we obtain
\begin{equation}\label{eq_5field_EFT_Delta1}
    \Delta_{\rm min}(Q_x,Q_y)=\frac{m_d}{\sqrt{c}}\sqrt{Q_x^2+Q_y^2}+\ldots\,.
\end{equation}
We can extract $c$ near four dimensions matching this result for $Q_x=Q_y$ with~\eqref{eq_5field_Delta_large}. We obtain
\begin{equation}
    c=\frac{1}{2}-\frac{\epsilon }{24}+O\left(\epsilon^2\right)\,.
\end{equation}

To evaluate the first correction to~\eqref{eq_5field_EFT_Delta1}, we compute the Casimir energy of the fluctuations. Fermions do not contribute at order $\log (Q)$ and thus we neglect them.  Expanding in fluctuations around the solution~\eqref{eq_5field_EFT_saddle} we find the following three bosonic modes
\begin{equation}\label{eq_5fields_lightmodes}
R\,\omega_{B,k=1,2,3}(\ell)=\begin{cases}
\ell \,,\\
\ell+d-2\,,\\
\sqrt{\ell(\ell+d-2)}\,.
\end{cases}
\end{equation}
Notice that the dispersion relations do not depend on $c$ nor the ratio $Q_x/Q_y$. From now on we set $d=4$ as appropriate to this order. The first two modes in~\eqref{eq_5fields_lightmodes} are as in~\eqref{eq_EFT_2modes} and do not contribute to the Casimir energy in dimensional regularization. The last mode is more interesting. Its contribution to the Casimir energy is logarithmically divergent in $d=4$
\begin{equation}\label{eq_1_loop}
\delta\Delta_{\rm min}^{(one-loop)} \supset \frac{R}{2}\sum_{modes}\sum_{\ell}n_{B}(\ell)\omega_{B,3}(\ell)\sim 
-\frac{1}{16}\sum_{\ell}\frac{1}{\ell}+\text{power-divergences}+\text{finite}
\,,
\end{equation}
where we used $n_{B,\ell}=(1+\ell)^2$ for the multiplicity in $d=4$. Cutting off the sum at $M\sim \sqrt{\epsilon (Q_x^2+Q_y^2)^{1/2}}/R$ and introducing counterterms to cancel the power-divergences as dictated by conformal invariance, we infer 
\begin{equation}\label{eq_5field_EFT_Casimir}
    \delta\Delta_{\rm min}^{(one-loop)}=-\frac{1}{32}\log \left(\epsilon \sqrt{Q_x^2+Q_y^2}\right)+O\left(Q_{x/y}^0\right)\,.
\end{equation}
The same result can be obtained evaluating the sum in dimensional regularization in a way that manifestly preserves the Weyl invariance of the EFT as in~\cite{Cuomo:2020rgt}. Note that the precise function of $Q_x/Q_y$ appearing in the logarithm in~\eqref{eq_5field_EFT_Casimir} depends on the $O(Q^0)$ higher derivative terms that we did not study. For $Q_x=Q_y$, \eqref{eq_5field_EFT_Casimir} agrees with the result~\eqref{eq_5field_Delta_large}, therefore providing a nontrivial check of the EFT methodology. Additionally, the EFT result in~\eqref{eq_5field_EFT_Casimir} predicts that the coefficient multiplying the logarithm is the same for all values of the ratio $Q_x/Q_y$.\footnote{We can also use the EFT to predict the $Q^0$ Casimir energy directly in $d=3$.  Since the fermionic dispersion relations for the light modes are independent of $c$,
\begin{equation}
\omega_{F,1}^{(\pm)}(\ell)=\omega_{F,2}^{(\pm)}(\ell)=p_F(\ell)\pm m_d\,,\qquad
\omega_{F,3}^{(\pm)}(\ell)=p_F(\ell)\,,
\end{equation}
fermions do not form a Fermi sphere and thus do not contribute the Casimir energy. We conclude
\begin{equation}
 \delta\Delta_{\rm min}^{(one-loop)}\vert_{d=3}\simeq -0.13255 \,.
\end{equation}
Note that the Casimir energy is negative for any value of $c$ and $Q_x/Q_y$.
\label{footnote_5fields_Casimir}}

\subsection{SQED}\label{sec:SQED}

$3d$ $\mathcal{N}=1$ SQED consists of a $U(1)$ gauge multiplet with $N_f$ fundamental matter multiplets and vanishing Chern-Simons level. As discussed in appendix \ref{app:epsilon_exp}, the theory possesses a $\mathbb{Z}_2$ R-symmetry which protects its moduli space. This moduli space is $2N_f-1$-dimensional moduli space, and is parametrized by the matter fields with one combination gauged away. The effective theory on the moduli space is a $\mathbb{CP}^{N_f-1}$ NLSM, consisting of $2N_f-2$ Goldstone bosons which together with the dilaton give the $2N_f-1$ massless bosonic degrees of freedom that were expected. As formerly mentioned, SQED with $N_f=2$ is thought to be dual to the 7-field WZ model discussed in section~\ref{sec:7_field}. We will study the theory directly in $3d$ using the EFT methods discussed in section \ref{sec:EFT_and_CFT}.

Since the effective theory consists of a dilaton coupled to a $\mathbb{CP}^{N_f-1}$ NLSM, the leading 2-derivative term in the bosonic part of the EFT takes the form (momentarily working in arbitrary spacetime dimensions)
\begin{equation}\label{eq_CPN_EFT}
    \mathcal{L}=\frac{1}{2}(\partial_\mu\Phi)^2+\frac{1}{2}m^2_d\Phi^2+2c \Phi^2 \gamma_{ij}\partial_\mu z_i\partial^\mu \bar{z}_{ j}\;.
\end{equation}
Here, $\Phi$ is the dilaton 
and the complex coordinates $z_i$, $i=1,...,N_f-1$ parametrize $\mathbb{CP}^{N_f-1}$. 
$\gamma_{ij}$ is the Fubini-Studi metric
\begin{equation}
    \gamma_{i{j}}=\frac{(1+|z|^2)\delta_{i j}-\bar{z}_i z_{ j}}{(1+|z|^2)^2} \;.
\end{equation}
Finally, $c$ is some constant which cannot be fixed from symmetry considerations. We have fixed its normalization such that at $c=1$ and $N_f=2$, the model reduces to three real free fields. 

We can focus on the specific $U(1)$ isometry rotating $z_1$ by a phase and compute $\Dmin(Q)$. Following the analysis in section \ref{sec:EFT_and_CFT}, we find a solution to the EOMs on the cylinder of the form 
\begin{equation}
    \Phi=v\;,\qquad z_1=e^{i\mu t}\;,
\end{equation}
with all other fields vanishing. The equations of motion set $\mu^2=\frac{m^2_d}{c}$, and the solution has charge and energy density
\begin{equation}
    J_0=cv^{2}\mu\;,\qquad T_{00}=m^2_d v^2\;.
\end{equation}
We thus find at the two-derivative level
\begin{equation}\label{eq_DQ_leading_SQED}
    \Dmin(Q)=\frac{m_dQ}{\sqrt{c}}+O(Q^0)\stackrel{d=3}{=}
    \frac{Q}{2\sqrt c}+O(Q^0)\;.
\end{equation}
As a consistency check, for $c=1$ we obtain the result for a free theory. We can also set $d=4-\epsilon$ and compare this calculation with the result for the 7 field WZ model. Matching with equation \eqref{eq_uRu_Delta}, we extract $c$ near four dimensions:
\begin{equation}\label{eq_5field_EFT_c}
c=1-\frac{1}{3}\epsilon+O\left(\epsilon^2\right)\,.
\end{equation}

\begin{figure}
    \centering
    \includegraphics[width=0.5\linewidth]{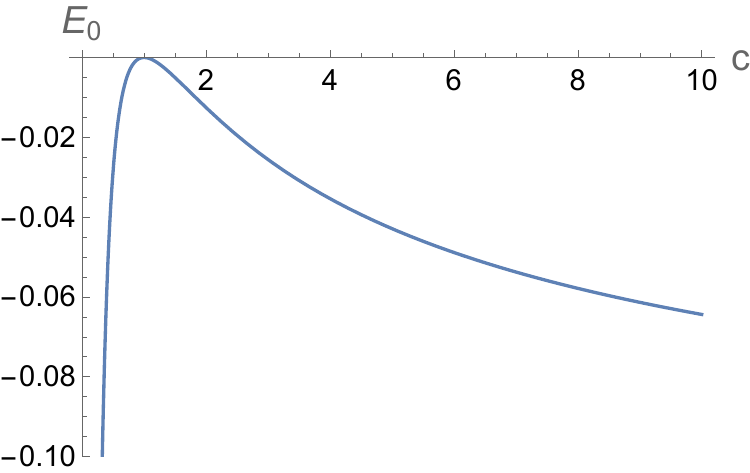}
    \caption{Casimir energy $E_0$ for the $\mathbb{CP}^1$ model.}
    \label{fig:CP1}
\end{figure}

The subleading term in $\Dmin(Q)$ will be the Casimir energy $E_0$, which will be a $c$-dependent constant. We compute it in appendix \ref{app:CP1casimir} for the physical number of dimensions, $d=3$. For the case $N_f=2$ where $\mathbb{CP}^1$ is just the 2-sphere, the result is displayed in figure \ref{fig:CP1}.
At $c=1$ the theory becomes free and the Casimir energy vanishes, as expected, while for all other values of $c$ it is negative. For the case $N_f>2$, the result is $E_0(N_f)=E_0(N_f=2)+2(N_f-2)e_0$ where $e_0$ is a non-positive $c$-dependent quantity described in appendix \ref{app:general_nf}. We thus find that fixing a $U(1)$ symmetry, the minimal-dimension operators with a fixed charge $Q$ should have dimension
\begin{equation}
    \Dmin(Q)=\frac{1}{2\sqrt c} Q+E_0(N_f)+O(1/Q)\;.
\end{equation}

\subsection{Approximate moduli in non-supersymmetric theories}

\label{sec:largeN}

While exact moduli spaces are hard to construct in interacting theories without supersymmetry, there are several models that admit flat directions in certain parametric limit $\delta\rightarrow 0$, where generically $\delta^2\sim 1/N$, while remaining interacting -- see e.g.~\cite{Rabinovici:1987tf,PhysRevLett.52.1188,Chai:2020zgq,Chai:2020onq}. From the viewpoint of the EFT~\eqref{eq_EFT}, this means that the coefficient of the dilaton potential~\eqref{eq_cc_dilaton} is parametrically suppressed with respect to its natural value.\footnote{For a theory with a unique quartic coupling $\sim \lambda$,  which may possibly be $O(1)$, on a solution~\eqref{eq_modulis_vev} a generic potential contributes as $M^d/\lambda$ to the energy, where $M$ is the generic gap of the massive modes on the moduli space.} Equivalently, the parameter $\delta$ induces a large separation between the (tachyonic) mass of the dilaton and the gap of the other massive modes in the moduli space, ensuring that states with spontaneously broken conformal symmetry are parametrically long lived.

It is interesting to repeat the EFT analysis of section~\ref{sec:EFT_and_CFT} in this setup. To this aim we simply include the potential~\eqref{eq_cc_dilaton} with a small coefficient $\sim\delta^{\frac{4}{d-2}}$ (where the powers have been chosen for convenience, such that $\delta$ would scale as a cubic coupling around a conformal vacuum).  Assuming that the gap of the massive modes in the would-be moduli space scales as $\sim g^2 v^2$ for some coupling $g$ (that in the most convenient normalization appears as an overall $1/g^2$ in front of the action and is possibly $O(1)$ for a strongly coupled model) as in the former examples, we find that the dilaton is light and the large charge operator spectrum obeys a linear relation as in~\eqref{eq_Deltamin_1} for $1\ll g^2 Q\ll 1/\delta^{2}$. However for $g^2 Q\gg 1/\delta^{2}$ eventually the dilaton and the other moduli acquire a gap of order $\delta^2 g^2 Q$, the only remaining light modes being the Goldstones for the internal symmetry; consequently, the scaling dimension grows nonlinearly with the charge $\Delta_{\rm min}(Q)\sim(\delta^2g^2 Q)^{\frac{d}{d-1}}/(\delta^2g^2)$, in agreement with the prediction of the large charge expansion \cite{Hellerman:2015nra,Monin:2016jmo}. In summary
\begin{equation}
\Delta_{\rm min}(Q)\propto\begin{cases}\displaystyle
Q &\text{for } 1\ll g^2 Q\ll 1/\delta^{2} \\
\displaystyle
\frac{(\delta^2g^2 Q)^{\frac{d}{d-1}}}{\delta^2g^2}
&\text{for } g^2 Q\gg 1/\delta^{2}\,.
\end{cases}
\end{equation}

The simplest model displaying this behaviour~\cite{Chai:2020zgq} is a scalar $O(N)\times O(N)$ CFT in $2<d<4$,   with potential given by
\begin{equation}\label{eq_ON_ON}
V(\vec{\phi}_1^{\,2},\vec{\phi}_2^{\,2})=\frac{\lambda_{11}}{8N}
(\vec{\phi}_1^{\,2})^2+\frac{\lambda_{22}}{8N}
(\vec{\phi}_2^{\,2})^2+\frac{\lambda_{12}}{4N}
\vec{\phi}_1^{\,2}\,\vec{\phi}_2^{\,2}\,,
\end{equation}
where we normalized the couplings so that they admit a natural large $N$-limit. To leading order in $N$, the beta functions of the coupling admits a fixed-point (for any $d$) such that
\begin{equation}\label{eq_ON_ON_lambda}
\lambda_{11}=\lambda_{22}=-\lambda_{12}\equiv\lambda>0\quad\text{for}\;N\rightarrow\infty\,,
\end{equation}
where $\lambda=16\pi^2\epsilon$ for $d=4-\epsilon$.
The potential~\eqref{eq_ON_ON} becomes a perfect square at such fixed points and the theory admits a flat direction for
\begin{equation}\label{eq_ON_ON_vacua}
\langle\vec{\phi}_1^{\,2}\rangle=\langle\vec{\phi}_2^{\,2}\rangle\,.
\end{equation}
In the vacua~\eqref{eq_ON_ON_vacua} the light modes consist of a dilaton and a $O(N)\times O(N)/O(N)$ NLSM, while the radial modes acquire a mass $\sim \lambda\langle\vec{\phi}^{\,2}_{1/2}\rangle/N$. Higher order corrections destroy the moduli space and contribute to a potential of the form
\begin{equation}
\frac{\lambda'}{4 N^2}\vec{\phi}_1^{\,2}\vec{\phi}_2^{\,2}\,.
\end{equation}
where $\lambda'=8\lambda$ at leading order in the $\epsilon$-expansion. This term induces a tachyonic mass $M_{dilaton}^2\sim M^2/N\ll M^2$ for the dilaton.

We may then apply the former discussion to this example, with $g^2\sim \epsilon/N$ and $\delta^2\sim 1/N$ (where $\epsilon$ could be $O(1)$). We conclude that operators in large traceless symmetric representations with $Q$ boxes under both the $O(N)$'s have approximately linear in $Q$ scaling dimension for $1\lesssim\epsilon Q/N\ll N$. Note that this prediction is not a trivial consequence of large $N$ factorization, since we are considering charges $Q\gtrsim N$. The behavior eventually changes for $\epsilon Q/N\gtrsim N$, for which we recover the nonlinear superfluid behavior. These predictions are simple to verify in the large charge double-scaling limit in the $\epsilon$-expansion (and it should be possible to compute the scaling dimension at large $N$ with $Q/N$ fixed).  We spare the details to the reader.

Charged operators in this model should be contrasted to the spectrum of large charge operators in the $O(N)$ model, studied in the $\epsilon$-expansion in \cite{Antipin:2020abu} and at large $N$ for any $d$ in \cite{Alvarez-Gaume:2019biu,Giombi:2020enj}. In that case there is no approximate moduli space, and the only relevant parameter is $\epsilon Q/N$, whose value controls the transition from the perturbative diagrammatic regime to the superfluid phase where $\Delta_{\rm min}(Q)\sim \frac{N}{\epsilon} (\epsilon Q/N)^{\frac{d}{d-1}}$, with no intermediate linear regime.

Another potentially interesting testing ground for our predictions could be the conformal Fishnet model \cite{Gurdogan:2015csr}, a non-supersymmetric and non-unitary deformation of $\mathcal{N}=4$ SYM with gauge group $SU(N_c)$ which at large $N_c$ becomes integrable.
It was recently suggested that in the planar limit this model admits flat directions with a spontaneously broken $U(1)$ symmetry~\cite{Karananas:2019fox}.\footnote{This was verified to one-loop order in the 't Hooft coupling.} Our EFT analysis of large charge operators does not rely in any essential form on unitarity and should therefore apply to operators with charge $Q/N_c\gg 1$ if the moduli space is indeed robust in the planar limit.\footnote{Encouragingly,  charged operators in the planar limit have been studied in \cite{Basso:2018agi}, where it was found that $\Delta_{\rm min}(Q)\simeq f(g) Q$ where $f$ is some nontrivial function of the strong coupling.  It remains unclear at this stage if this predictions survives for $Q\gtrsim N_c$.} It would also be interesting to explore connections between the large charge double-scaling limit and integrability in that model, along the lines of~\cite{Caetano:2023zwe}.

We finally mention that our discussion provides a simple necessary condition for the existence of a dilaton at the endpoint of the conformal window of QCD \cite{Ellis:1970yd,Appelquist:2010gy}. The existence of a flat direction in the conformal window of QCD would be a rather finely tuned phenomenon, as emphasized in \cite{Gorbenko:2018ncu,Zan:2019idm,Benini:2019dfy}, but it is nonetheless conceivable at large $N_c$, as in the example above. If such a scenario is indeed realized, it is natural to expect that also the internal $SU(N_f)\times SU(N_f)$ symmetry is spontaneously broken to the diagonal in the moduli space, as in the gapped phase. It would then follow from our discussion that at the endpoint of the conformal window the lowest dimensional operators with charge $Q$ under a Cartan generator in the diagonal of $SU(N_f)\times SU(N_f)$ (corresponding to traceless symmetric representations with $2Q$ boxes in both Young diagrams) would have scaling dimension $\Delta_{min}(Q)\propto Q$ for $1\ll Q\ll N_c $. We hope that in the future this prediction can be used as a test of the existence of a flat direction at the endpoint of the QCD conformal window.

\subsection{Convexity in CFTs}

Motivated by the weak gravity conjecture, the paper \cite{Aharony:2021mpc} conjectured a convexity relation for $\Dmin(Q)$ in unitary $d>2$ CFTs. Specifically, the charge convexity conjecture (CCC) states that there exists some $q_0$ of order 1 for which
\begin{equation}\label{eq:convexity_conj}
    \Dmin(n_1q_0+n_2q_0)\geq \Dmin(n_1q_0)+\Dmin(n_2q_0)\;,\qquad \forall \;n_1,n_2\,\in \mathbb{N}\;.
\end{equation}
Although a counterexample was found in \cite{Sharon:2023drx}, the conjecture still seems to apply to a large range of CFTs, and an obvious question is whether some alternative version of it still holds.\footnote{One such version is studied in \cite{Aharony:2023ike}.} It is natural to start from the weakest version, where we allow any (arbitrarily large) $q_0$. We call this version of the conjecture the weak CCC. 

Originally the weak CCC was dismissed as trivial, since all known phases at asymptotically large charge were manifestly convex in the sense of \eqref{eq:convexity_conj}.
However, in the models that we discussed this is no longer true. Specifically, in our examples $\Dmin(Q)$ is given in an expansion of the form 
\begin{equation}\label{eq:3d_expansion}
    \Dmin(Q)=\alpha_1Q+\alpha_0+O(1/Q)
\end{equation}
in $3d$, or
\begin{equation}
    \Dmin(Q)=\alpha_1Q+\alpha_0\log Q+ O(Q^0)
\end{equation}
in the $\epsilon$-expansion from $4d$. In both cases convexity is not manifest, and instead the sign of the Casimir energy $\alpha_0$ determines whether the weak CCC holds (if $\alpha_0$ vanishes, one needs to compute higher-order terms to check the weak CCC, although we have not encountered such examples). 
It is then interesting to ask whether the weak CCC is indeed obeyed in these examples.

Happily, in all of the examples discussed above $\alpha_0$ is non-positive (and is only zero in free theories or for BPS operators in theories with extended supersymmetry), and so the weak CCC holds. In particular, in the $5$-field WZ model and the SQED example discussed in sections \ref{subsec_5_field} (see footnote \ref{footnote_5fields_Casimir}) and \ref{sec:SQED}, the Casimir energy is always negative, and so even though we cannot fix the Wilson coefficient $c$ we are still assured that convexity holds. 

We comment that convexity is a misnomer in the $3d$ case, since convexity for a function $f(x)$ requires $f''(x)\geq 0$, while the condition \eqref{eq:convexity_conj} instead corresponds to superadditivity. This distinction was not important in previous discussions of convexity since these two definitions were equivalent in all of the examples that were considered (see \cite{Aharony:2021mpc,Antipin:2021rsh,Ishii:2024pnh}). However,  in the expansion \eqref{eq:3d_expansion} $\alpha_0$ determines superadditivity while higher-order terms determine convexity, and so the two concepts are distinct. We will keep calling the condition \eqref{eq:convexity_conj} ``convexity'' to avoid confusion.

One might wonder if the Casimir energy~\eqref{eq_EFT_Casimir} is always non-positive in $d=3$, perhaps assuming $\mathcal{N}=1$ supersymmetry. We were not able to establish this result in full generality. However we found that the contributions of scalars with \emph{relativistic} dispersion relations~\eqref{eq_EFT_n2_modes_relativistic} to the Casimir energy is always negative for $d=3$ and $d=4$.\footnote{This is not true anymore in $d=5$ and $d=6$, where the sign depends on the mass $n_k$ in~\eqref{eq_EFT_n2_modes_relativistic}.} Additionally, as already mentioned in section~\ref{sec:EFT_and_CFT}, fermions always contribute negatively to the Casimir energy.  These observations are enough to establish that $\alpha_0\leq 0$ in a large class models, including all the ones that we studied here. Although this is a limited set of examples, it is reassuring that they all obey the weak CCC.   It would be nice to prove the weak CCC in general.\footnote{The weak CCC can be proven in two dimensions \cite{Palti:2022unw}. } 

Note however that it is easy to construct moduli EFTs that violate the weak CCC, e.g. the theory of an axio-dilaton (or its $3d$ $\mathcal{N}=1$ version) with the appropriate sign of the higher derivative terms in~\eqref{eq_EFT_4derivatives} or a purely bosonic $\mathds{R}\times\mathbb{CP }^1$ model (with the dilaton potential~\eqref{eq_cc_dilaton} finely tuned to zero) as in~\eqref{eq_CPN_EFT}, with $c\gtrsim 4.63$ in $d=5$ or with $0<c<2$ in $d=6$, so that the Casimir energy is negative. Therefore a proof of the weak CCC, if possible,  must go beyond EFT arguments (e.g. of the form appearing in \cite{Orlando:2023ljh}) and must rely nontrivially on the requirement that the theory admits a well behaved UV completion.

\section{Large charge macroscopic limit for theories with moduli spaces}\label{sec:macroscopic}

Given the relation between moduli spaces and large charge operators that we discussed so far,
we expect that at distances much smaller than the sphere radius, correlation functions in a large charge state reduce to correlators in the moduli space. This statement in fact follows from EFT at distances $R\gg |x|\gg 1/m$, where $m\propto Q^{\frac{1}{d-2}}/R$ is the EFT cutoff, in this section we further assume that it remains true also at distances $|x|\ll 1/m$.\footnote{More precisely, correlation  functions are expected to be well approximated by a nontrivial flat space limit up to corrections of order $|x|^2/R^2$ and $1/(mR)^2\sim Q^{-\frac{2}{d-2}}$; in practice we often do not know how to compute these corrections beyond the EFT regime, and therefore we do not discuss them here.} These considerations relate the scaling dimension spectrum  of large charge operators and the mass spectrum of particles on the moduli space.
 This set of ideas is formalized by the existence of a macroscopic limit \cite{Lashkari:2016vgj}, first introduced in the context of charged operators in \cite{Jafferis:2017zna}. In the context of SCFTs with moduli spaces, this limit was further explored in~\cite{Caetano:2023zwe}. For the benefit of the reader, here we provide a self-contained discussion.

\subsection{Macroscopic limits}

Let us begin by reviewing the main ideas behind the concept of macroscopic limit, following~\cite{Jafferis:2017zna}.
Let us consider a CFT with a $U(1)$ symmetry (the generalization to more complicated groups being straightforward), but not necessarily a moduli space 
for the moment. Let us denote with $\mO_Q$ the lowest dimensional operator as a function of the charge for each allowed value of $Q$; additional specifications might be needed when this is not unique. The macroscopic limit conists in sending $Q\rightarrow\infty$ and $R\rightarrow\infty$ with the ratio $Q^{\frac{1}{\gamma}}/R=\text{fixed}$, where the power $\gamma$ depends on the theory and the state, such that correlation functions of light operators (whose dimensions are held fixed in the limit) between two large charge states remain finite. The limit must be taken so that the dimensionful distances between the light operators remain finite. For instance, consider the expectation value of a light scalar operator $\mO_{\delta,q}$ of dimension $\delta\sim O(1)$ and charge $q\sim O(1)$ in a large charge state
\begin{equation}
\langle Q |\mO_{\delta,q}|Q+q\rangle=\frac{\lambda_{Q,\mO,Q+q}}{R^{\delta}}\,,
\end{equation}
In the macroscopic limit this matrix element should reduce to an appropriate expectation value in a nontrivial flat-space state:
\begin{equation}\label{eq_OPE_macro_requirement}
\langle Q |\mO_{\delta,q}|Q+q\rangle    
\xrightarrow{\text{macro.}}
\langle \text{flat}| \mO_{\delta,q}|\text{flat}\rangle_{\mathds{R}^d}
\end{equation}
where we indicated with $|\text{flat}\rangle$ the resulting state, that is characterized by the dimensionful scale $\nu\equiv Q^{\frac{1}{\gamma}}/R$ - therefore dimensional analysis implies $\langle \text{flat}| \mO_{\delta,q}|\text{flat}\rangle_{\mathds{R}^d}\propto \nu^{\delta}$.
Using that the macroscopic limit leaves the ratio $Q^{\frac{1}{\gamma}}/R=\text{fixed}$, \eqref{eq_OPE_macro_requirement} can be stated purely in terms of CFT data as:
\begin{equation}\label{eq_OPE_macro}
\lim_{Q\rightarrow\infty}Q^{-\delta/\gamma}\lambda_{Q,\mO,Q+q}=\nu^{-\delta}
\langle \text{flat}| \mO_{\delta,q}|\text{flat}\rangle_{\mathds{R}^d}=
\text{finite}\,.
\end{equation}
As a less trivial example, consider the correlator of two light operators in the large charge ground-state
\begin{equation}\label{eq_macro_4pt}
\langle Q|\mO_{\delta_1,q_1}(\tau,\hat{n}_1)\mO_{\delta_2,q_2}(0,\hat{n}_2)|Q+q_1+q_2\rangle\xrightarrow{\text{macro.}}\underbrace{\langle \text{flat}|\mO_{\delta_1,q_1}(\tau,\vec{x})\mO_{\delta_2,q_2}(0,\vec{0})|\text{flat}\rangle_{\mathds{R}^d}}_{
\equiv g_{\mO\mO}(\tau,|\vec{x}|)}\,,
\end{equation}
where we work on the Euclidean cylinder with metric  $ds^2=d\tau^2+R^2 d\hat{n}^2$ and the limit is taken keeping fixed $\tau$ and $\vec{x}^{\,2}=2R^2(1-\hat{n}_1\cdot\hat{n}_2)$. Introducing cross ratios in standard notation
\begin{equation}
    \langle\mO_Q(0)\mO_{\delta_1,q_1}(z,\bar{z})\mO_{\delta_2,q_2}(1)\mO_{-Q}(\infty)\rangle\equiv G_Q(z,\bar{z})\,,
\end{equation}
~\eqref{eq_macro_4pt} can be stated purely in terms of the flat space four-point function as
\begin{equation}
\lim_{Q\rightarrow\infty}Q^{-\frac{\delta_1+\delta_2}{\gamma}}G_Q\left(z=1-\nu\frac{w}{Q^{\frac{1}{\gamma}}},\bar{z}=1-\nu\frac{\bar{w}}{Q^{\frac{1}{\gamma}}}\right)=\nu^{-\frac{\delta_1+\delta_2}{\gamma}}\times g_{\mO\mO}(\text{Re}w,|\text{Im}w|)\,,
\end{equation}
where the factors of $\nu$ ensure that $w$ has the dimensions of a length. Note that for the limit to be nontrivial it is crucial that the result is nonzero at least for some correlators.

The existence of a macroscopic limit is a natural requirement; physically it corresponds to the statement that zooming in at distances much shorter than the sphere radius in a very large charge state we recover the dynamics of a nontrivial flat space state. However, to identify such state and the power $\gamma$ in the fixed ratio $Q^{1/\gamma}/R$ we need to make some assumptions. Below we discuss some possibilities.

Suppose that the scaling dimensions of the operators $\mO_Q$ grows as $\Delta(Q)\sim Q^{\alpha}$ for $Q\rightarrow\infty$. We additionally assume that the corresponding state on the sphere is approximately homogeneous, and therefore the spin carried by these operators does not scale with $Q$.  The energy density $\varepsilon$ and the charge density $\rho$ of the corresponding state therefore are
\begin{equation}
\varepsilon\sim \frac{Q^{\alpha}}{R^d}\,,\qquad
\rho\sim\frac{Q}{R^{d-1}}\,.
\end{equation}
In the simplest option, the macroscopic limit is defined so that the energy density remains fixed in the limit $R\rightarrow\infty$ \cite{Lashkari:2016vgj}. This implies that the ratio $Q/R^{\frac{d}{\alpha}}$ must be held fixed,  i.e. $\gamma=\frac{d}{\alpha}$. We have three possibilities:
\begin{enumerate}
    \item for $\alpha>\frac{d}{d-1}$ the charge density vanishes; we expect that in this case the flat limit corresponds to a generic high energy state, of the kind analyzed in \cite{Komargodski:2021zzy}, and that the operators $\mO_Q$ do not provide the large charge ground state.
    \item for $\alpha=\frac{d}{d-1}$ we obtain the relation $\varepsilon\sim \rho^{\frac{d}{d-1}}$; this is the only possible equation of state for a CFT at finite density in flat space and corresponds to the expected behaviour of large charge operators in generic CFTs.
    \item for $\alpha<\frac{d}{d-1}$ we find that $\rho$ diverges, contradicting the assumption that correlation functions remain finite in the limit. We thus conclude that we cannot take the macroscopic limit while keeping the energy density fixed. 
\end{enumerate}
This is a simple but important conclusion: if $\Delta_{\rm min}(Q)\sim Q^{\alpha}$ with $\alpha<\frac{d}{d-1}$, then the existence of a nontrivial macroscopic limit necessarily results in a nontrivial state with vanishing energy density, and therefore the theory must have a flat direction. More in general, we can drop the assumption that the scaling dimensions grows as a power law in $Q$ and we conclude:
\begin{equation}\label{eq_scaling_modulus}
\lim_{Q\rightarrow\infty}\Delta_{\rm min}(Q)/ Q^{\frac{d}{d-1}}=0\;\cup\;
\exists\text{ macro. limit }Q\rightarrow\infty\quad\implies\quad\exists \text{ moduli space}\,.
\end{equation}
When the energy density vanishes in the macroscopic limit, we expect that  $\rho$ vanishes as well - otherwise we would have vacua that break spontaneously the Lorentz symmetry. This implies that $\gamma<d-1$.

\subsection{Applications}

The conclusion~\eqref{eq_scaling_modulus} is a powerful one, and it essentially argues for the opposite direction of the arrow in the argument~\eqref{eq_moduli_to_Delta}. However at this stage it does not tell us what is the right value of $\gamma$. We expect that, when there exist a family operators satisfying the property~\eqref{eq_scaling_modulus}, the $U(1)$ symmetry is spontaneously broken in the flat space state. In this case we can resort to the EFT analysis of the previous sections to identify the proper way to take the macroscopic limit. From the solution~\eqref{eq_cyl_sol} we see that the physical limit corresponds to holding the dilaton's vev $v$ fixed. Therefore, using the expression for the charge density~\eqref{eq_Q_fix} we consider the limit
\begin{equation}\label{eq_macro_Q}
Q\rightarrow\infty\,,\quad  R\rightarrow\infty\,,\quad\text{with}\quad\frac{Q}{R^{d-2}}=\text{fixed}\,,
\end{equation}
which amounts at $\gamma=d-2$. To be more precise, the fixed ratio $Q/R^{d-2}$ is related to the dilaton's vev in the broken vacuum by~\eqref{eq_v_Q}. In this section we discuss some consequences of the existence of the limit~\eqref{eq_macro_Q} for correlation functions. For simplicity, we also assume that the lowest dimensional charged operators $\mO_Q$ are scalars.

Let us warm-up by considering OPE coefficients. Setting $\gamma=d-2$, \eqref{eq_OPE_macro} implies that OPE coefficients of scalar operators scale as
\begin{equation}\label{eq_macro_OPE_scalar}
\lambda_{Q,\mO,Q+q}\sim Q^{\Delta_{\mO}/(d-2)}\,,
\end{equation}
for large $Q$. This is indeed the result that we obtained within EFT \eqref{eq_EFT_OPE}. Similarly, the requirement that the macroscopic limit reproduces the moduli space theory implies that OPEs of spinning operators should vanish faster than in~\eqref{eq_macro_OPE_scalar} for $Q\rightarrow\infty$ - this is again trivially realized in the EFT, see~\eqref{eq_EFT_OPE_spin}.

In some cases, the existence of a finite macroscopic limit follows from supersymmetry. As a trivial example, consider a rank $1$ $4d$ $\mathcal{N}=2$ SCFT, and denote with $\Phi_r$ the $U(1)_r$ charge $r$ Coulomb branch operator, whose scaling dimension is $\Delta_r=r$. Working in the standard normalization where the chiral ring's OPE is unit normalized, supersymmetry fixes the form of the so called extremal correlators, consisting of $n+1$ chiral operators and one anti-chiral operator (see e.g. \cite{Eberhardt:2020cxo}):
\begin{equation}\label{eq_extremal}
\langle\Phi_Q(0)\Phi_{r_1}(x_1)\Phi_{r_2}(x_2)\ldots
\Phi_{r_n}(x_n)\bar{\Phi}_{Q+\sum_i r_i}(x)\rangle=
\frac{w(Q+\sum_i r_i)}{|x|^{2\Delta_Q}}\prod_i\frac{1}{|x_i-x|^{2\Delta_i}}\,,
\end{equation}
where
\begin{equation}
w(r)=\langle\Phi_{r}(0)\bar{\Phi}_{r}(1)\rangle\,.
\end{equation}
Mapping the correlator~\eqref{eq_extremal} to the cylinder we obtain
\begin{equation}
\langle Q|\Phi_{r_1}(\tau_1,\hat{n}_1)\Phi_{r_2}(\tau_2,\hat{n}_2)\ldots
\Phi_{r_n}(\tau_n,\hat{n}_n)|Q\rangle=\sqrt{\frac{w(Q+\sum_i r_i)}{w(Q)}}\prod_i\frac{e^{\tau_i\Delta_{i}/R}}{R^{\Delta_i}}\,.
\end{equation}
Taking the macroscopic limit and using the localization result~\cite{Grassi:2019txd,Beccaria:2020azj} $\sqrt{w(Q+\sum_i r_i)/w(Q)}\simeq Q^{\sum_i r_i/2}$, we obtain
\begin{equation}\label{eq_macro_extremal}
\langle Q|\Phi_{r_1}(\tau_1,\hat{n}_1)\Phi_{r_2}(\tau_2,\hat{n}_2)\ldots
\Phi_{r_n}(\tau_n,\hat{n}_n)|Q\rangle\xrightarrow{macro}
v^{\sum_i\Delta_i}\,,
\end{equation}
where we identified the fixed-ratio $v\equiv \sqrt{Q}/R$ with the local coordinate $v$ on the Coulomb branch. We see that taking the macroscopic limit in~\eqref{eq_macro_extremal} we obtained a coordinate-indepent result.  This matches the well known holomorphic properties of chiral operators in the Coulomb branch.

The macroscopic limit implies a relation between the particles' mass spectrum in the moduli space and the dimensions of charged primary operators that are much heavier than the large charge ground state. The exact correspondence can be established relating the conformal block decomposition of a four-point function with two insertions of the large charge operator $\mO_Q$ and two light operators, with the K\"allen Lehman decomposition in the moduli space.  We discuss this relation below. It is important that the macroscopic limit applies at arbitrary short distances, even if the EFT breaks for $\vec{x}^{\,2},\tau^2\ll m^{-2}\sim R^2/Q^{\frac{2}{d-2}} $. For technical reasons that we mention below, our discussion will be limited to $d\geq 4$.

Consider the two point function of a light neutral operator $\mO$ (whose $O(1)$ quantum numbers will be irrelevant for our purposes) in the large charge ground state:
\begin{equation}\label{eq_4pt_light_light}
F_{Q,\mO}(\tau,\hat{n}_1\cdot\hat{n}_2)\equiv\langle Q|\mO(\tau,\hat{n}_1)\mO(0,\hat{n}_2)|Q\rangle
\,.
\end{equation}
It is convenient to evaluate this correlator inserting a complete set of states between the two light operators. Grouping states into conformal families, this results into the conformal block decompostion. For $\tau>0$ we use the s-channel conformal block decomposition 
\begin{equation}\label{eq_4pt_light_light_OPE}
F_{Q,\mO}(\tau,\hat{n}_1\cdot\hat{n}_2)=\frac{1}{R^{2\Delta_{\mO}}}\left[\lambda_{Q,\mO,Q}^2 g_{\Delta_Q,0}^{\mO_Q,\mO}(\tau,\hat{n}_2\cdot\hat{n}_1)+\sum_{\Delta,\ell}
\lambda_{Q,\mO,(\Delta,\ell)}^2
g_{\Delta,\ell}^{\mO_Q,\mO}(\tau,\hat{n}_2\cdot\hat{n}_1)\right]\,,
\end{equation}
where we isolated the contribution of the operator $\mO_Q$ itself and $g_{\Delta,\ell}^{\mO_Q,\mO}$ is the conformal block corresponding to the exchange of an operator of dimension $\Delta$ and spin $\ell$ in the $\mO_Q$-$\mO$ OPE.  We normalized the s-channel conformal blocks so that the primary's contribution reads:
\begin{equation}\label{eq_CBs}
g_{\Delta,\ell}^{\mO_Q,\mO}(\tau,\hat{n}_2\cdot\hat{n}_1)= e^{-\tau(\Delta-\Delta_Q)/R}C_{\ell}^{\left(\frac{d}{2}-1\right)}(\hat{n}_1\cdot\hat{n}_2)+\text{descendants}\,.
\end{equation}
The s-channel decomposition in these coordinates converges for $\tau>0$. For $\tau<0$ we may use the t-channel expansion, which in the case of identical neutral operators $\mO$ is obtained from the s-channel decomposition with the replacement $\tau\rightarrow-\tau$ in the blocks.

In the macroscopic limit~\eqref{eq_macro_4pt}, the correlation function~\eqref{eq_4pt_light_light} reduces to the moduli space correlator. This in turn can be expressed using the spectral decomposition, i.e.
\begin{equation}\label{eq_macro_4pt_moduli}
\begin{split}
F_{Q,\mO}(\tau,\hat{n}_1\cdot\hat{n}_2)&\xrightarrow{\text{macro.}}\langle \mO(\tau,\vec{x})\mO(0,\vec{0})\rangle_{v,\vec{\phi}} \\
&=\langle\mO\rangle_{v,\vec\phi}^2+\int_0^{\infty} dm^2\rho_{\mO\mO}(m^2) G_{m^2}(\tau^2+\vec{x}^{\,2})\,,
\end{split}
\end{equation}
where $\rho_{ij}$ is the spectral density
\begin{equation}\label{eq_spectral_density}
\rho_{ij}(p^2)=(2\pi)^{d-1}\sum_{n}\langle 0|\mO_i|n\rangle\langle n|\mO_j|0\rangle\delta^d(q_n-p)\,.
\end{equation}
We would like to relate~\eqref{eq_macro_4pt_moduli} with the conformal block decomposition described above. To this aim we need to compute the form of the conformal blocks in the macroscopic limit. 

The macroscopic limit of the conformal block depends on whether the gap $\Delta-\Delta_Q$ and the angular momentum $\ell$ of the exchanged operator scale or not with $R$. When $\Delta-\Delta_Q\sim \ell\sim O(1)$, the conformal blocks are trivial in the macroscopic limit
\begin{equation}
g_{\Delta,\ell}^{\mO_Q,\mO}(\tau,\hat{n}_2\cdot\hat{n}_1)\xrightarrow{\text{macro.}}
1\qquad\text{for }\;\Delta-\Delta_Q\sim \ell\sim O(1)\,.
\end{equation} 
This result arises because the contributions of descendants in~\eqref{eq_CBs} is suppressed by powers of $\Delta_Q$, see e.g. \cite{Komargodski:2021zzy}. Therefore operators with $O(1)$ gap and angular momentum can only contribute to the disconnected part of the correlator in the macroscopic limit. The contribution is nontrivial only when the corresponding OPE coefficient scales as $R^{\Delta_{\mO}}\sim Q^{\frac{\Delta_{\mO}}{d-2}}$. Because of~\eqref{eq_OPE_macro_requirement}, we find immediately that the exchange of $\mO_Q$ in~\eqref{eq_4pt_light_light_OPE} reproduces exactly the disconnected term in~\eqref{eq_macro_4pt_moduli} in the macroscopic limit.  Therefore we conclude that the OPE coefficients of all the other exchanged operators vanish in the macroscopic limit
\begin{equation}\label{eq_macro_OPE_other}
\lim_{Q\rightarrow\infty}Q^{-\frac{\Delta_{\mO}}{d-2}}    \lambda_{Q,\mO,(\Delta,\ell)}=0\,.
\end{equation}

The connected term in~\eqref{eq_macro_4pt_moduli} arises from the exchange of operators with $\Delta-\Delta_Q\sim R$ and/or angular momentum $\ell\sim R$. We therefore set
\begin{equation}
    \omega\equiv\frac{\Delta-\Delta_Q}{R}\,,\qquad
    k\equiv\frac{\ell}{R}\,,\qquad\bar{v}\equiv\frac{\Delta_Q}{R^{d-2}}\,.
\end{equation}
The discussion of the macroscopic limit of the conformal blocks at fixed $\omega$ and $k$ is similar to the one in \cite{Komargodski:2021zzy}, where the authors considered the macroscopic limit at fixed energy density. The contribution of the \emph{n}th level descendants in~\eqref{eq_CBs} scales as $\sim \left[(\omega^2+k^2)R^2/\Delta_Q\right]^n\sim R^{n[2-(d-2)]}$. Descendants may therefore be neglected for $d>4$. In $d=4$ we may instead resort to the exact expression of the conformal block \cite{Dolan:2000ut,Dolan:2003hv}, and we find that descendants only contribute to an overall prefactor. We spare the details to the reader, and only report the final result:
\begin{equation}\label{eq_CBs_2}
g_{\Delta,\ell}^{\mO_Q,\mO}(\tau,\hat{n}_2\cdot\hat{n}_1)\xrightarrow{\text{macro.}}
\mathcal{N}_{d,\bar{v}}(\omega,k)\frac{ R^{d-3}}{k}e^{-\omega\tau}\int d^{d-1}p\,\delta(|\vec{p}|-k)e^{i\vec{k}\cdot\vec{x}}\,,
\end{equation}
where the prefactor reads
\begin{equation}\label{eq_CBs_2_prefactor}
\mathcal{N}_{d,\bar{v}}(\omega,k)=\frac{R^{d-3}}{2^{d-2}\pi^{\frac{d-2}{2}}\Gamma\left(\frac{d-2}{2}\right)}
\times\begin{cases}
\displaystyle 1 &\text{for }d>4\\
\displaystyle \exp\left(\frac{\omega^2+k^2}{2\bar{v}^2}\right) &\text{for }d=4\,.
    \end{cases}
\end{equation}
We expect that a result analogous to~\eqref{eq_CBs_2} holds also in $d<4$ with a different prefactor, but we did not prove it.

Using~\eqref{eq_CBs_2}, we recast the macroscopic limit of~\eqref{eq_4pt_light_light_OPE} as
\begin{equation}\label{eq_4pt_light_light_macro_final}
F_{Q,\mO}(\tau,\hat{n}_1\cdot\hat{n}_2)\xrightarrow{\text{macro}.}  
\frac{\lambda_{Q,\mO,Q}^2}{R^{2\Delta_{\mO}}}+\int_0^\infty d\omega\int \frac{d^{d-1}k}{(2\pi)^{d-1}}e^{-\omega|\tau|}e^{i\vec{k}\cdot\vec{x}}K(\omega,|\vec{k}|)\,,
\end{equation}
where we defined
\begin{equation}\label{eq_macro_K_def}
 K(\omega,k)= (2\pi)^{d-1}\sum_{\Delta,\ell}\delta\left(\omega-\frac{\Delta-\Delta_Q}{R}\right) 
 \delta\left(k-\frac{\ell}{R}\right) \frac{\mathcal{N}_{d,\bar{v}}(\omega,k)\lambda^2_{Q,\mO,(\Delta,\ell)}}{k\,R^{2\Delta_{\mO}+3-d}}\,.
\end{equation}
The existence of the macroscopic limit therefore requires that the combination in~\eqref{eq_macro_K_def} becomes $R$-independent in the appropriate sense.\footnote{To this aim, it might be necessary to smoothen the delta functions in~\eqref{eq_macro_K_def} with normalized averages over intervals of infinitesimal size $\delta$ around $\omega$ and $k$ - see e.g. footnote $8$ in \cite{Komargodski:2021zzy}.}

Finally, comparing~\eqref{eq_4pt_light_light_macro_final} with~\eqref{eq_macro_4pt_moduli}, we arrive at
\begin{equation}\label{eq_macro_spectral}
    \theta(\omega)\rho_{\mO\mO}(\omega^2-k^2)=K(\omega,k)\,.
\end{equation}
\eqref{eq_macro_spectral} establishes a direct connection between the flat space spectral density and the operators exchanged in the s- and t- channel of the correlation function~\eqref{eq_4pt_light_light}.

As a simple application,  consider the contribution of a mass $m$ particle in the spectral density in $d=4$:
\begin{equation}\label{eq_rho_massive_macro}
\theta(\omega)\rho_{\mO\mO}(\omega^2-k^2)=c_{m^2}\bar{v}^{2\Delta_{\mO}-2}\theta(\omega)\delta(\omega^2-k^2-m^2)\,,
\end{equation}
where the powers of $\bar{v}$ are dictated by dimensional analysis and $c_{m^2}$ is an arbitrary dimensionless coefficient. Because of~\eqref{eq_macro_spectral},  \eqref{eq_rho_massive_macro} implies that there exists one or more operators with scaling dimension
\begin{equation}\label{eq_macro_Delta_massive}
\Delta-\Delta_Q\simeq\sqrt{\ell^2+R^2m^2}\,,
\end{equation}
and with OPE coefficient
\begin{equation}\label{eq_macro_OPE_massive}
\begin{split}
\sum \lambda^2_{Q,\mO,(\Delta,\ell)}&\simeq c_{m^2}(R\bar{v})^{2\Delta_{\mO}-2}\frac{4\pi\ell}{\sqrt{\ell^2+R^2m^2}}
e^{-\frac{R^2 m^2+2\ell^2}{2\Delta_Q}}\\
&\propto Q^{\Delta_{\mO}-1}\frac{\ell}{\sqrt{\ell^2+R^2m^2}}e^{-\frac{R^2 m^2+2\ell^2}{2\Delta_Q}}
\,,
\end{split}
\end{equation}
where the sum runs over all the degenerate operators with the same scaling dimension~\eqref{eq_macro_Delta_massive}. Note that it is important that the mass remains finite in the macroscopic limit $R m\sim Q^{1/2}$.

\eqref{eq_macro_Delta_massive} and~\eqref{eq_macro_OPE_massive} are in agreement with the explicit results of \cite{Caetano:2023zwe}, where the authors studied correlation functions with two large charge Coulomb branch operators in $\mathcal{N}=4$ SYM with gauge group $SU(2)$ in the weak coupling double-scaling limit and conjectured a similar formula.\footnote{To compare the results, note that $e^{-\frac{\omega^2+k^2}{2\bar{v}^2}}\simeq 1+O(g^2)$ at weak coupling in the double-scaling limit.} We also expect the prediction~\eqref{eq_macro_Delta_massive} to be relevant in large $N$ SCFTs with moduli spaces.

Let us conclude this dicussion with a comment about the macroscopic limit for $\mathcal{N}=4$ SYM in the holographic limit as explained in \cite{Ivanovskiy:2024vel}.  We focus for concreteness on the Higgsing pattern $SU(N)\rightarrow U(1)\times SU(N-1)$. Holographically this is realized by displacing a brane at some distance from the center of the bulk of AdS$_5$ \cite{Klebanov:1999tb}. The state which reduces to this vacuum in the macroscopic limit is the one created by a fully symmetric chiral operator $\sim \text{Tr}[X^Q]$ with $Q\sim N$, which is dual to a giant graviton spherical $D3$ brane in AdS$_5$ \cite{Hashimoto:2000zp} (see also \cite{McGreevy:2000cw,Balasubramanian:2001nh,Corley:2001zk,Lin:2004nb}). For large charge $Q\gg N$ (but still $Q\ll N^2$ in order to neglect the gravitational backreaction) the brane extends towards the boundary of AdS similarly to the moduli space one. Zooming in at distances much smaller than the $S^3$ radius the two branes are locally identical. The mapping between moduli space particles and charged CFT operators then follows straightforwardly. In particular, for both branes we can consider open strings ending on them. For the moduli space brane the string state could be e.g. a W-boson particle, whose mass $T\Delta X\sim g v$ is proportional to the distance between the displaced brane and the bulk of AdS.  For the giant graviton with $Q\gg N$ the string corresponds to a W-boson like primary operator,  and the semiclassical energy of the string provides the scaling dimensions in agreement with~\eqref{eq_macro_Delta_massive}.\footnote{We thank Shota Komatsu for clarifying discussions about this point.} It would be interesting to further explore  the connection between large charge operators and moduli spaces in holography.

Let us conclude by mentioning that it might be interesting to apply the macroscopic limit to the study of partition functions at finite chemical potential, and in particular to the supersymmetric index. For instance, in $4d$ $\mathcal{N}=2$ SCFTs certain limits of the supersymmetric index are related to the physics of the Coulomb and Higgs branches of the theory (see~e.g.~\cite{Rastelli:2016tbz, Rastelli:2014jja} for reviews); we expect that many of these relations can be understood by a generalization of the arguments presented in this section.

\section{Comments on moduli spaces with no broken charges}\label{sec:comments}

The main conceptual message of this work could be summarized as follows. In CFTs with moduli spaces where a global charge is also broken, large charge operators create, in radial quantization, \emph{semiclassical} states that closely resemble the vacua where both the conformal and the internal symmetry are broken. The connection is clear in the EFT saddle-point~\eqref{eq_cyl_sol}, since the dilaton acquires a large value. This is what allows to prove  the existence of a tower of operators whose scaling dimension grows  linearly with the charge. The existence of a conserved charge is essential, as one can identify a precise tower of states, those corresponding to the lowest dimensional charged operators. 

One might still wonder whether
in the general case (where
no internal symmetry is spontaneously broken)
the existence of a moduli space
is still  reflected in {\it some}
semiclassical  feature of the spectrum.
Below we address this question starting from the EFT of a real dilaton. Not surprisingly, we conclude that in the absence of a conserved charge it is not possible to identify a tower of special operators that look like the spontaneously broken  vacua. However, we speculate that the moduli space is reflected in the existence of certain resonant states on the cylinder, whose width becomes parametrically narrower than their energy in the high energy limit.

The basic idea is simple. Consider for concreteness the EFT for a real dilaton, which is relevant in the $ABC$ model \cite{Gaiotto:2018yjh} and other $3d$ $\mathcal{N}=1$ SCFTs. The leading term is simply the free action
\begin{equation}\label{eq_dilaton_real_EFT}
S=\int d^dx\left[\frac{1}{2}(\pd\Phi)^2-\frac{m_d^2}{2}\Phi^2+\ldots\right]\,,
\end{equation}
where the dots stand for higher derivative terms and additional fields, such as the superpartners. 
On the cylinder, 
the free theory admits a homogeneous solution of the form
\begin{equation}\label{eq_phi_cyl}
\Phi=\frac{v^{\frac{d-2}{2}}}{\sqrt{2}}\cos( m_d t+t_0)\,.
\end{equation}
The parameter $t_0\sim t_0+2\pi$ is a zero-mode of the solution and should be integrated over in the path-integral.
In appendix~\ref{eq_phi_cyl} we show that the solution~\eqref{eq_phi_cyl} arises as the saddle-point describing the two-point function $\langle \Phi^n(x_i)\Phi^n(x_f)\rangle$ for $n\gg 1$ in free theory, with
\begin{equation}
v^{d-2}=\frac{n}{(d-2)\Omega_{d-1}R^{d-2}}\,.
\end{equation}
Therefore the solution~\eqref{eq_phi_cyl} describes a primary state with dimension $\Delta_n=n$ in free theory. Note that from the cylinder viewpoint, the quantization condition $n\in\mathds{N}$ follows from the semiclassical quantization of the field around the saddle-point.\footnote{The simplest way to see this is to consider the Bohr-Sommerfeld condition
\begin{equation}
 \oint \Pi d\Phi   =2\pi\,,
\end{equation}
where $\Pi=R^{d-1}\int d^{d-1}\hat{n}\,\dot{\Phi}$ is the field momentum and the integration is over an oscillation period $T=2\pi/m_d$ of the solution.}

One might therefore naively conclude from the existence of the solution~\eqref{eq_phi_cyl} that CFTs with moduli spaces always admit states with energy $\Delta_n\simeq n$ for $n\gg 1$.\footnote{Note that this would be a stronger claim than the one for charged operators, since the ratio $\Delta_n/n$ is fixed by the EFT. This is because there is no free parameter in the EFT action (up to rescalings of the dilaton) unlike the charged case.} There is an important subtlety however. Even when $n\gg 1$ the dilaton vev does not stay large for the full oscillation period $T=2\pi/m_d$ of the solution. This is to be contrasted with the large charge profile~\eqref{eq_cyl_sol}, where the dilaton retains a constant value and the solution never exits the EFT valdity regime. 

Within EFT, we can trust the profile~\eqref{eq_phi_cyl} only as long as higher derivative terms are suppressed, i.e. when
\begin{equation}
    (\pd\Phi)^2\ll|\Phi|^{\frac{2d}{d-2}}\,.
\end{equation}
From this condition we infer that for each oscillation period the solution~\eqref{eq_phi_cyl} exits the EFT validity regime for a time $\delta t$ of the order
\begin{equation}\label{eq_deltat_T}
    \frac{\delta t}{T}\sim \left(\frac{m_d}{v}\right)^{\frac{d-2}{d}}\sim n^{-\frac{1}{d}}\,.
\end{equation}
Physically, during the time in which the dilaton vev is small, all the degrees freedom of the CFT become again massless and interact nontrivially with the dilaton mode. The breakdown of EFT therefore signifies that the EFT Hamiltonian is not complete, and there exist additional off-diagonal terms that mix the dilaton with the other fields of the theory. These terms are suppressed by $\delta t/T\sim n^{-1/d}$ with respect to the leading contribution $\Delta_n\simeq n$ at large energy, but they imply nonetheless that we cannot diagonalize the Hamiltonian restricting ourselves only to the EFT states. Therefore the energy eigenstates are complicated linear combinations of dilaton quasi-particles and other strongly coupled modes. This was in some sense a foreordained conclusion:
the spectrum of a general CFT is dense at high energies, and it is a priori not clear how to identify special eigenstates.

This discussion is again to be contrasted with the case of charged operators - in which case the saddle-point~\eqref{eq_cyl_sol} describes the ground state in a certain fixed charge sector. In that case the EFT is obtained integrating out only massive states in a standard fashion, and we can self-consistently focus on the EFT Hilbert space. In the neutral case instead, when using the EFT around the solution~\eqref{eq_phi_cyl} we are neglecting other degenerate states with the same quantum numbers, that are not separated by a large gap. There is no selection rule that prevents the semiclassical dilaton state~\eqref{eq_cyl_sol} from mixing with these other states made of strongly coupled degrees of freedom.

Yet, there remain the question of which is the right interpretation of the semiclassical state described by~\eqref{eq_phi_cyl}. After all, this is certainly a well defined state for short enough times, such that the dilaton did not yet have time to interact significantly with other modes. Our speculation is that the saddle-point~\eqref{eq_phi_cyl} describes a resonant state. Indeed, in general the spectrum of states of energy $\Delta \gg 1$ is expected to be exponentially dense \cite{Bhattacharyya:2007vs,Shaghoulian:2015lcn,Benjamin:2023qsc} and may therefore be approximated by a continuum.\footnote{For a detailed discussions of the validity of this approximations, see e.g.  \cite{Giudice:2000av,Cuomo:2020fsb}.} On general grounds, the mixing of an isolated state with a continuum results in the existence of a resonant pole in the Green's function of the theory \cite{di1992lezioni}. Therefore we expect that the interactions between the dilaton and other strongly coupled modes will result into a width that controls the lifetime of the state~\eqref{eq_phi_cyl}. We can estimate the width on dimensional grounds from the ratio~\eqref{eq_deltat_T}
\begin{equation}\label{eq_width}
\Gamma_n\sim \Delta_n\times\frac{\delta t}{T}\sim n^{\frac{d-1}{d}}\,.
\end{equation}

We can also reach this conclusion using EFT. Indeed, it is sometimes the case that in EFT one can integrate out a continuum of states at the price of losing unitarity. The most famous example is perhaps the EFT for $CP$ violation in Kaon physics, in which one describes the physics in a small energy window around the Kaon mass and neglects the pions and other light states, whose effect is described via the inclusion of small imaginary terms in the Hamiltonian. In that case the EFT does not describe the full evolution of the system, but only the probability that the Kaon evolves into another quasi-degenerate particle rather than decaying into ligher ones.

Assuming that is is possible to integrate out the quasi-continuum of degenerate states around the solution~\eqref{eq_phi_cyl}, we need to parametrize in EFT our inability to describe the interactions of the dilaton with the other modes during the small interval of times in which the dilaton vev becomes small. Similarly to~\cite{Hellerman:2020eff}, as long as $\delta t/T\ll 1$ this can be done treating the interaction as instantaneous, introducing effective operators which are localized in field space $\propto\delta(\Phi)$. The leading such operator is
\begin{equation}\label{eq_dS_real}
\delta S=c\int d^dx\,\delta(\Phi)\left[(\pd\Phi)^4+\ldots\right]^{\frac12+\frac{d-2}{4 d}}\,,
\end{equation}
where the powers are dictated by Weyl invariance and the dots stand for terms which are needed to Weyl complete the parenthesis.\footnote{In the notation of section~\ref{sec:moduliEFT}, the term in parenthesis can be written as
\begin{equation}
\frac{(d-2)^2}{4d(d-1)}\hat{\mathcal{R}}_{\mu\nu}^2\Phi^{4\left(1+\frac{2}{d-2}\right)}=(\pd\Phi)^4+\ldots  \,.  
\end{equation}} The general idea is that $(\pd\Phi)^{\frac{2}{d}}\sim \left[m_d^2 v^{d-2}\right]^{\frac{1}{d}}\gg 1/R $  controls the derivative expansion at the singular points where $\Phi=0$; on physical grounds, the reason why we can still use EFT is that the interactions with the strongly coupled modes is almost instantaneous, and should therefore be describable by effective \emph{source terms} localized at $m_d t+t_0\approx \frac{\pi}{2}+\pi\mathds{Z}$. \eqref{eq_dS_real} realizes the leading such term in a manifestly Weyl invariant fashion.

On the saddle-point, the operator~\eqref{eq_dS_real} results into a contribution to the stress tensor of the form
\begin{equation}
T_{00}\propto c\,
\delta(m_d t+t_0)|\dot{\Phi}|^{\frac{1}{2}+\frac{d-2}{2d}}
\simeq c\times \delta(m_d t+t_0) \left[m_d v^{\frac{d-2}{2}}\right]^{\frac12+\frac{d-2}{2d}}
\,.
\end{equation}
Thus, upon averaging over the zero-mode $t_0$, we obtain a correction $\sim c \,n^{\frac{d-1}{d}}$ to the energy. Since the operator~\eqref{eq_dS_real} has zero support in the usual moduli space, the coefficient $c$ is unknown and may be taken complex according to our discussion. This is therefore the leading contribution to the width in agreement with the estimate~\eqref{eq_width}. 

Matrix elements in such resonant states behave very differently than in generic high energy states. Indeed, we can compute the expectation value of an arbitrary scalar operator $\mO_{\delta}$ of dimension $\delta \sim O(1)$ via operator matching $\mO_{\delta}\propto |\Phi|^\delta$ as in section~\ref{sec:correlation_functions}. We therefore find that $t=0$ matrix elements take the form
\begin{equation}\label{eq_res_mat}
\langle v| \mO_{\delta}(t=0)|v\rangle\propto\Delta^{\frac{\delta}{d-2}}\,.
\end{equation}
Which is very different than the expected result in a generic (thermalized) high energy state $\langle \Delta|\mO_{\delta}(t=0)|\Delta\rangle\propto  \Delta^{\frac{\delta}{d}}$. These considerations suggest these resonances should be more properly interpreted as \emph{scar states}, similarly to \cite{Milekhin:2023was}.

To summarize, in the absence of a conserved charge it does not seem possible to find a specific tower of primary operators that carries over the information of the moduli space physics. It might however still be possible to construct long-lived resonances which, in a sense, can be thought as homogeneous excitations of the dilaton.

\section*{Acknowledgments}
We thank Zohar Komargodski and Shota Komatsu
for useful discussions.  The work of GC was supported by the Simons Foundation grant 994296 (Simons Collaboration on Confinement and QCD Strings). 
The work of LR was supported in part by the National Science Foundation under Grant NSF PHY-2210533 and by the  Simons Foundation under Grant 681267 (Simons Investigator Award).

\appendix

\section{Fluctuations around the large charge ground state}\label{app:fluctuations}

We expand the moduli around the solution~\eqref{eq_cyl_sol}
\begin{equation}
\Phi=v^{\frac{d-2}{2}}+\delta\Phi\,,\quad
\pi^a=m^a t/R+\delta \pi^a/v^{\frac{d-2}{2}}\,,\quad\phi^A=\bar{\phi}^A+\delta \phi^A/v^{\frac{d-2}{2}}\,.
\end{equation}
Expanding~\eqref{eq_EFT_cyl}, we find the following quadratic action
\begin{equation}
\begin{split}
\mL_{fluct}&=\Gb_{aB}\pd_\mu\delta\pi^a\pd^\mu\delta\phi^B
+\frac{m^a}{R}\dot{\delta\phi^B}\left(\Gb_{aB,C}\delta\phi^C
+2\Gb_{aB}\delta\Phi\right)
+\frac{m^a}{R}\dot{\delta\pi^b}\left(\Gb_{ab,A}\delta\phi^A
+2\Gb_{ab}\delta\Phi\right)
\\
&
+\frac12\Gb_{ab}\pd^\mu\delta\pi^a\pd_\mu\delta\pi^b
+\frac{1}{2R^2} m^a m^b \left(\Gb_{ab}\delta\Phi^2+2\Gb_{ab,C}\delta\Phi\delta\phi^C+\frac12\Gb_{ab,CD}\delta\phi^C\delta\phi^D\right)
\\
&+\frac12\Gb_{AB}\pd^\mu\delta\phi^A\pd_\mu\delta\phi^B
-\frac12\left(\Gb_{\Phi\Phi}\,\delta\Phi+\Gb_{\Phi\Phi,A}\delta\phi^A\right)
\pd^2\delta\Phi
\\
&-\frac{m_d^2}{2}\left(\Gb_{\Phi\Phi}\,\delta\Phi^2+2\Gb_{\Phi\Phi,A}\delta\Phi\delta\phi^A+\frac{1}{2}\Gb_{\Phi\Phi,AB}\delta\phi^A\delta\phi^B\right)
\,,
\end{split}
\end{equation}
where $\Gb=\hat{G}(\bar{\phi})$ and the additional subscripts denote derivatives with respect to the $\phi^A$'s, e.g. $\Gb_{ab,C}=\pd \hat{G}_{ab}/\pd\phi^C\vert_{\phi=\bar{\phi}}$, $\Gb_{\Phi\Phi,A}=\pd \hat{G}_{\Phi\Phi}/\pd\phi^A\vert_{\phi=\bar{\phi}}$, etc. . Using the EOMs~\eqref{eq_PHI_eom} and~\eqref{eq_phi_EOM}, we see that the terms proportional to $\delta\Phi^2$ and $\delta\Phi\delta\phi^C$ cancel; therefore, integrating by parts the term proportional to $\pd^2\delta\Phi$, we arrive at
\begin{equation}
\begin{split}
\mL_{fluct}&=\frac12\Gb_{ab}\pd^\mu\delta\pi^a\pd_\mu\delta\pi^b+\frac{m^a}{R}\dot{\delta\pi^b}\left(\Gb_{ab,A}\delta\phi^A
+2\Gb_{ab}\delta\Phi\right)
\\
&
+\frac{1}{2}\bar{G}_{\Phi\Phi}\,\pd_\mu\delta\Phi\pd^\mu\delta\Phi+\frac12\bar{G}_{\Phi\Phi,A}\pd_\mu\delta\phi^A\pd^\mu\delta\Phi
+
\frac12\Gb_{AB}\pd^\mu\delta\phi^A\pd_\mu\delta\phi^B
-\frac{1}{2} M^2_{AB}\delta\phi^A\delta\phi^B
\\
&+ \Gb_{aB}\pd_\mu\delta\pi^a\pd^\mu\delta\phi^B+\frac{m^a}{R}\dot{\delta\phi^B}\left(\Gb_{aB,C}\delta\phi^C
+2\Gb_{aB}\delta\Phi\right)\,,
\end{split}
\end{equation}
where we defined
\begin{equation}
M^2_{AB}= \left(\frac{m_d^2}{2}\bar{G}_{\Phi\Phi,AB}-\frac{m^c m^d}{2R^2}\,\Gb_{cd,AB}\right)\,.
\end{equation}
At this point it is convenient to isolate in the $\delta\pi$'s the component parallel to the chemical potential vector, according to the scalar product defined by $\Gb_{ab}$. We thus decompose
\begin{equation}
\delta\pi^a=m^a\delta\pi_0+\delta\pi^a_{\bot}\quad\text{such that}\quad
m^a\bar{G}_{ab}\delta\pi^b_{\bot}=0\,.
\end{equation}
Note that this decomposition is trivial when there is a unique spontaneously broken Cartan generator.  Using~\eqref{eq_PHI_eom} and~\eqref{eq_phi_EOM}, we recast the action as
\begin{equation}\label{eq_app_fluct_action_pre}
\begin{split}
\mL_{fluct}
&=\frac{1}{2}\Gb_{\Phi\Phi}\,\left(\pd^\mu\delta\Phi+\frac12\frac{\Gb_{\Phi\Phi,A}}{\Gb_{\Phi\Phi}}\pd^\mu\delta\phi^A\right)^2+R^2m_d^2\frac{\Gb_{\Phi\Phi}}{2}\left(\pd_\mu\delta\pi_0+ \frac{m^a\Gb_{aB}}{R^2 m_d^2\Gb_{\Phi\Phi}}\pd^\mu\delta\phi^B\right)^2
\\
&
+2\Gb_{\Phi\Phi}\,R m_d^2\,\left(\dot{\delta\pi_0}
+ \frac{m^a\Gb_{aB}}{R^2 m_d^2\Gb_{\Phi\Phi}}\dot{\delta\phi^B}\right)
\left(\delta\Phi+\frac12\frac{\Gb_{\Phi\Phi,A}}{\Gb_{\Phi\Phi}}\delta\phi^A
\right)
\\&
+\frac12\Gb_{ab}\pd^\mu\delta\pi^a_{\bot}\pd_\mu\delta\pi^b_{\bot}
+ \Gb_{aB}\pd_\mu\delta\pi^a_{\bot}\pd^\mu\delta\phi^B
+\frac{m^a}{R}\dot{\delta\pi^b_{\bot}}\Gb_{ab,A}\delta\phi^A
\\&
+\frac12\left(\Gb_{AB}-\frac{\Gb_{\Phi\Phi,A}\Gb_{\Phi\Phi,B}}{4\Gb_{\Phi\Phi}}
- \frac{m^c\Gb_{cA}m^d\Gb_{dB}}{R^2 m_d^2\Gb_{\Phi\Phi}}
\right)\pd^\mu\delta\phi^A\pd_\mu\delta\phi^B
\\&
+\left(\frac{m^a}{R}\Gb_{aB,C}-\frac{ m^a\Gb_{aB}\bar{c}_C}{R\bar{c}}\right)\dot{\delta\phi^B}\delta\phi^C
-\frac{1}{2} M^2_{AB}\delta\phi^A\delta\phi^B\,.
\end{split}
\end{equation}
We can rewrite~\eqref{eq_app_fluct_action_pre} compactly by defining
\begin{equation}
K^2_{AB}=\Gb_{AB}-\frac{\Gb_{\Phi\Phi,A}\Gb_{\Phi\Phi,B}}{4\Gb_{\Phi\Phi}}
- \frac{m^c\Gb_{cA}m^d\Gb_{dB}}{R^2 m_d^2\Gb_{\Phi\Phi}}\,,\qquad
d_{BC}=\frac{m^a}{R}\Gb_{aB,C}-\frac{ m^a\Gb_{aB}\Gb_{\Phi\Phi,C}}{R\Gb_{\Phi\Phi}}\,.
\end{equation}
Then, shifting $\delta\Phi$ and $\delta\pi_0$ as
\begin{equation}
\delta\Phi\rightarrow\delta\Phi-\frac12\frac{\Gb_{\Phi\Phi,A}}{\Gb_{\Phi\Phi}}\delta\phi^A\,,\qquad
\delta\pi_0\rightarrow\delta\pi_0-\frac{m^a\Gb_{aB}}{R^2 m_d^2\Gb_{\Phi\Phi}}\delta\phi^B\,,
\end{equation}
we arrive at
\begin{equation}\label{eq_app_fluct_action}
\begin{split}
\mL_{fluct}
&=\frac{1}{2}\Gb_{\Phi\Phi}\left[\left(\pd\delta\Phi\right)^2+R^2m_d^2(\pd\delta\pi_0)^2+4\,R m_d^2\,\dot{\delta\pi_0}\delta\Phi
\right]
\\&
+\frac12\Gb_{ab}\pd^\mu\delta\pi^a_{\bot}\pd_\mu\delta\pi^b_{\bot}
+ \Gb_{aB}\pd_\mu\delta\pi^a_{\bot}\pd^\mu\delta\phi^B
+\frac12 K^2_{AB}\pd^\mu\delta\phi^A\pd_\mu\delta\phi^B
\\&
+\frac{m^a}{R}\Gb_{ab,A}\dot{\delta\pi^b_{\bot}}\delta\phi^A
+d_{AB}\dot{\delta\phi^A}\delta\phi^B
-\frac{1}{2} M^2_{AB}\delta\phi^A\delta\phi^B\,.
\end{split}
\end{equation}

Upon canonically normalizing the fields, we recognize that the first line of~\eqref{eq_app_fluct_action} coincides with the action for the fluctuations around the helical solution~\eqref{eq_free_th_sol} in free theory. We therefore immediately conclude that the corresponding dispersion relations are given by~\eqref{eq_EFT_2modes} in the main text.

The other $n-2$ fields are also associated  with a Fock spectrum of light quasi-particles.  For large angular momentum $\ell \gg 1$ we can drop the terms in the third line of~\eqref{eq_app_fluct_action} and we obtain~\eqref{eq_EFT_n2_modes_large_l} in the main text.

Due to the terms with a single time derivative, it is hard to compute in full generality the spectrum associated with the fields in the second and third line of~\eqref{eq_app_fluct_action}. A simplification occurs when the first and second term in the last line of~\eqref{eq_app_fluct_action} vanish. This happens for instance in theories in which there is a unique Cartan charge and a symmetry acting as $\pi\rightarrow-\pi$ on the Goldstone boson, e.g. a model with a single $U(1)$ charge invariant under charge conjugation. In this case the spectrum of fluctuations results into standard relativistic dispersion relations, cf.~\eqref{eq_EFT_n2_modes_relativistic}.

\section{Moduli spaces and \texorpdfstring{$\epsilon$}{epsilon}-expansion for \texorpdfstring{$3d$}{3d} \texorpdfstring{$\mathcal{N}=1$}{N=1} SUSY theories}\label{app:epsilon_exp}

Three-dimensional $\mathcal{N}=1$ SUSY has two supercharges, and as a result SUSY is much less powerful compared to the more extensively-studied cases with four and more supercharges. In particular, $3d$ $\mathcal{N}=1$ theories do not possess a continuous R-symmetry, and as a result some famous features which are attributed to SUSY like non-renormalization theorems and protected operator dimensions do not oocur. Nonetheless, a discrete R-symmetry is sometimes available, and can be enough to provide some exact results. One particular example involves a $\mathbb{Z}_2$ R-symmetry which in some cases protects a moduli space from being lifted to all orders in perturbation theory~\cite{Affleck:1982as,Gremm:1999su,Gukov:2002es}, which we now review following \cite{Gaiotto:2018yjh}. 

Interactions which lift moduli spaces in $3d$ $\mathcal{N}=1$ theories take the form of a superpotential\footnote{Our $3d$ $\mathcal{N}=1$ superspace conventions follow \cite{Gates:1983nr}. In particular we have bosonic coordinates $x^\mu$ and Majorana Grassmanian coordinates $\theta_\alpha$, and a basic scalar multiplet consists of a real scalar and Majorana fermion.} 
\begin{equation}
    \int d^2\theta W(\Phi_i)\;,
\end{equation}
with $\Phi_i$, $i=1,...,N$ some real scalar superfields. A $\mathbb{Z}_2$ R-symmetry acts as $d^2\theta\to -d^2\theta$, and so it appears only if one can assign a transformation $W\to -W$ to the superpotential under it. If this symmetry appears, it severely restricts the terms which can appear in $W$ once quantum corrections are taken into account. 

One immediate example of a $3d$ $\mathcal{N}=1$ theory with an exact moduli space is $\mathcal{N}=1$ SQED, consisting of a gauge multiplet coupled to $N_f$ complex scalar multiplets $\Phi_i$ of charge 1 and no Chern-Simons term. Classically the theory has a moduli space parametrized by the vev of $\Phi_i$ up to an overall phase rotation, so that the moduli space is $\mathbb{CP}^{N_f-1}$. The theory possesses a $\mathbb{Z}_2$ R-symmetry which acts on $\Phi_i$ and which prevents the appearance of any superpotential under quantum corrections. Indeed, gauge invariance and a global $SU(N_f)$ symmetry forces the superpotential to be a function of $|\Phi_i|^2$, but all such terms are even under the $\mathbb{Z}_2^R$ and so cannot be generated. The moduli space is thus exact quantum-mechanically, at least to all orders in perturbation theory. The absence of a Chern-Simons term is crucial here, since otherwise the $\mathbb{Z}_2$ R-symmetry is explicitly broken, although there are examples where an R-symmetry is emergent in the IR and which leads to a moduli space \cite{Choi:2018ohn,Aharony:2019mbc}). 

A particularly simple example is a WZ model which consists of three real scalar superfields and the superpotential 
\begin{equation}\label{eq_ABC}
    W=\frac{g}{2}ABC\;.
\end{equation}
At the classical level this theory has a moduli space where one of the three fields $A,B,C$ has a vev while the other do not. The theory also has a $\mathbb{Z}_2^R$ symmetry which acts as $A\to -A$, together with an $S_3$ permutation symmetry. The combination of these symmetries protects the moduli space to all orders in perturbation theory, since it constrains the effective action to include only terms of the form
\begin{equation}
    W_{\text{eff}}\supset ABCf(A^2,B^2,C^2)\;,
\end{equation}
for some function $f$.

\subsection{\texorpdfstring{$\epsilon$}{epsilon}-expansion for \texorpdfstring{$3d$ $\mathcal{N}=1$}{3d N=1}}\label{sec:epsilon_exp}

There is an immediate obstruction to studying  $3d$ $\mathcal{N}=1$ theories in the $\epsilon$-expansion from $4d$: $3d$ $\mathcal{N}=1$ theories have two supercharges, while $4d$ SUSY theories cannot have less than 4. A workaround for WZ theories was proposed in \cite{Fei:2016sgs,ThomasSeminar}. Consider a general $3d$ $\mathcal{N}=1$ WZ theory consisting of scalar multiplets $\Psi_i$ for $i=1,...,N_\Psi$. In components these consist of real Bosons $\phi_i$ and Majorana Fermions $\psi_i$, and the Lagrangian takes the form
\begin{equation}\label{eq:Lag_WZ}
    \mL=\frac12(\partial_\mu\phi_i)^2-\frac{i}2\psi_i^\alpha\partial_{\alpha\beta}\psi_i^\beta-\frac{i}{2}\partial_i\partial_jW\psi^\alpha_i\psi_{\alpha j}-\frac12(\partial_iW)^2\;,
\end{equation}
with $W(\Psi_i,\bar\Psi_i)$ the superpotential. 
To approach this theory in the $\epsilon$-expansion, we instead consider a generalization of the Lagrangian with an additional flavor index $a=1,...,N_f$ for the fermions:
\begin{equation}\label{eq_eps_ep}
     \mL=\frac12(\partial_\mu\phi_i)^2-\frac{i}2\psi_{ai}^{\alpha}\partial_{\alpha\beta}\psi_{ai}^\beta-\frac{i}{2}\partial_i\partial_jW\psi^\alpha_{ai}\psi_{\alpha a j }-\frac12(\partial_iW)^2\;.
\end{equation}
This Lagrangian is non-SUSY, except for the case $N_f=1$ which reduces to \eqref{eq:Lag_WZ}. The advantage of stduying this generalized theory is that for $N_f$ a multiple of 2, we can repackaging the $3d$ two-component Majorana fermions into $N_f/2$ $4d$ four-component Majorana fermions, which allows us to study the theory in the $\epsilon$-expansion from $4d$. In our case we work with $N_f$ a multiple of 4, and repackage the fermions into $N_f/4$ $4d$ Dirac Fermions. The corresponding Lagrangian in $4d$ is
\begin{equation}
\label{eq:4d_lagrangian1}
    \mL=\frac12(\partial_\mu\phi_i)^2+i\bar{\psi}_{bi}\slashed{\pd}\psi_{bi}-\partial_i\partial_jW\bar\psi_{bi}\psi_{ b j }-\frac12(\partial_iW)^2\;,
\end{equation}
where we sum over $b=1,...,N_f/4$. The Lagrangian \eqref{eq:4d_lagrangian1} can now be studied in the $\epsilon$-expansion as usual, and we can compute various CFT data as a function of $N_f$. Eventually we set $N_f=1$ which corresponds to the $3d$ $\mathcal{N}=1$ theory of interest. This trick has been tested for various $3d$ $\mathcal{N}=1$ theories \cite{Fei:2016sgs,Liendo:2021wpo,Benini:2018bhk,Benini:2018umh} and provided excellent agreement with other methods.

\section{One-loop scaling dimension in the \texorpdfstring{$5$}{5}-field WZ model}\label{app_5fields}

The Lorentzian action for the model~\eqref{eq_5field_W} on $\mathbb{R}\times S^{d-1}$ in $d=4-\epsilon$ dimensions reads:
\begin{equation}
\begin{split}
\mL&=
|\pd x|^2-m_d^2|x|^2+|\pd y|^2-m_d^2|y|^2+\frac12(\pd a)^2-\frac{m_d^2}{2}a^2
\\&
+i\bar{\tilde{a}}\slashed{\pd}\tilde{a}+
i\bar{\tilde{x}}\slashed{\pd}\tilde{x}
+i\bar{\tilde{x}}_c\slashed{\pd}\tilde{x}_c
+i\bar{\tilde{y}}\slashed{\pd}\tilde{y}
+i\bar{\tilde{y}}_c\slashed{\pd}\tilde{y}_c
-\frac{1}{4}g^2 a^2(|x|^2+|y|^2)-\frac{g^2}{8}(|x|^2-|y|^2)^2
\\&
-\frac{g}{2} a\left(\bar{\tilde{x}}\tilde{x}+\bar{\tilde{x}}_{c}\tilde{x}_c
-\bar{\tilde{y}}\tilde{y}-\bar{\tilde{y}}_{c}\tilde{y}_c
\right)
-\frac{g}{2} \left[\bar{\tilde{a}}\left(\tilde{x}_c x+\tilde{x}x^*-\tilde{y}_c y-\tilde{y} y^*\right)+c.c.\right]\,,
\end{split}
\end{equation}
where tilde's denote Dirac fields and we neglected the fermion multiplicity index. Note that to each of the charged scalars $x,\, y$ correspond two Dirac fields, $(\tilde{x},\tilde{x}_c)$ and $(\tilde{y},\tilde{y}_c)$, with opposite charge.

To study fluctuations around the saddle-point~\eqref{eq_5field_saddle}, we define $r_{x/y}$ and $\pi_{x/y}$ via
\begin{equation}
    x=\frac{(v_x+r_x)}{\sqrt{2}}e^{-i\mu_x t-i\pi_x}\,,\qquad
     y=\frac{(v_y+r_y)}{\sqrt{2}}e^{-i\mu_y t-i\pi_y}\,,
\end{equation}
and we make a field redefinitions to make the fermions neutral
\begin{equation}
\tilde{x} \rightarrow e^{-i \mu_x t-i \pi_x t}  \tilde{x}\,, \qquad
\tilde{y} \rightarrow e^{-i \mu_y t-i \pi_y t}  \tilde{y}\,, 
\qquad
\tilde{x}_c \rightarrow e^{+i \mu_x t+i \pi_x t}  \tilde{x}_c\,, \qquad
\tilde{y}_c \rightarrow e^{i \mu_y t+i \pi_y t}  \tilde{y}_c\,.
\end{equation}
Specializing to the solution~\eqref{eq_5field_saddle_equalQ}, we find the bosonic and fermionic quadratic actions:
\begin{align}\nonumber
    \mL^{(2)}_{bos} &=\sum_{a=x,y}\left[\frac{1}{2}(\pd r_a)^2+\frac{1}{2}(\pd \pi_a)^2+2 m_d r_a\dot{\pi}_a-\frac{1}{4}M^2r_a^2\right]-\frac{M^2}{2}r_xr_y+\frac12(\pd a)^2-\frac{m_d^2+M^2}{2}a^2 \\[0.4em]
\mL^{(2)}_{fer}&=
i\bar{\tilde{x}}\left(\slashed{\pd}-i\gamma^0 m_d\right)\tilde{x}
+i\bar{\tilde{x}}_c\left(\slashed{\pd}+i\gamma^0 m_d\right)\tilde{x}_c
+i\bar{\tilde{y}}\left(\slashed{\pd}-i\gamma^0 m_d\right)\tilde{y}
+i\bar{\tilde{y}}_c\left(\slashed{\pd}+i\gamma^0 m_d\right)\tilde{y}_c\nonumber
\\
&+i\bar{\tilde{a}}\slashed{\pd}\tilde{a}
-\frac{M}{2}\left[\bar{\tilde{a}}\left(\tilde{x}_c+\tilde{x}-\tilde{y}_c-\tilde{y}\right)+c.c.\right]\,,
\end{align}
where we defined 
\begin{equation}
    M=\frac{g |v_x|}{\sqrt{2}}\,.
\end{equation}

It is simple to obtain the spectrum of bosonic fluctuations. We find
\begin{equation}\label{eq_app_5field_omegaB}
\begin{aligned}
\omega_{B,1/2}(\ell)&=\sqrt{J_{\ell}^2+m_d^2}\mp m_d\,,\\
\omega_{B,3/4}(\ell)&=\sqrt{J_{\ell}^2+2 m_d^2+\frac{M^2}{2} \mp \sqrt{4 J_{\ell}^2 m_d^2+\left(2 m_d^2+\frac{M^2}{2}\right)^2}}\,, \\
\omega_{B,5}(\ell)&=\sqrt{J_{\ell}^2+m_d^2+ M^2}\,,
\end{aligned}
\end{equation}
where $J_{\ell}^2=\ell(\ell+d-2) / R^2$. Note that the first three dispersion relations coincide with the EFT result~\eqref{eq_5fields_lightmodes} when expanded for $M^2\gg J_{\ell}^2,1/R^2$, while $\omega_{B,4}$ and $\omega_{B,5}$ describe gapped modes.

To obtain the spectrum of fermions, it is convenient to proceed as in \cite{Sharon:2020mjs,Antipin:2022naw} and write the Dirac fields in terms of Weyl spinors. Expanding the latter in terms of spinor harmonics \cite{Camporesi:1995fb}, we find that the spectrum contains four modes with dispersion relation
\begin{equation}\label{eq_app_5field_omegaF_12}
    \omega^{(\pm)}_{F,1}(\ell)=\omega^{(\pm)}_{F,2}(\ell)=p_F(\ell)\pm m_d\,,
\end{equation}
where $p_F(\ell)=\left(\ell+\frac{d-1}{2}\right)/R$. The other six dispersion relations are given by the positive roots, each counted twice, of the following cubic equation for $\omega^2$:
\begin{multline}\label{eq_app_5field_cubic}
    \omega^6-
    \omega^4 \left[2 \left(M^2+m_d^2\right)+3 p_F^2(\ell)\right]
    +\omega^2 \left[
    3 p_F^4+4 M^2 p_F^2(\ell)+\left(M^2+m_d^2\right)^2
    \right]\\-
    p_F^2(\ell) \left[M^2-m_d^2+p_F^2(\ell )\right]^2=0\,.
\end{multline}
Since the polynomial in~\eqref{eq_app_5field_cubic} is itself the product of two real cubic polynomials in $\omega$ (identical up to the flip $\omega\rightarrow-\omega$) with positive discriminant, \eqref{eq_app_5field_cubic} admits three real and positive solutions for $\omega^2$. Their explicit form is complicated in general, but simplifies in certain limits. For instance, the expansion of the nontrivial fermionic dispersion relations for $R^2 M^2\ll 1$ is:
\begin{align}
\omega^{(\pm)}_{F,3}(\ell)& =p_{F}(\ell)+
\frac{2 p_{F}(\ell) M^2}{4 p_{F}^2(\ell)-m^2_d}
-\frac{2 p_{F}(\ell) M^4 \left[4 p_{F}(\ell)^2+m^2_d\right]}{\left[4 p_{F}^2(\ell)-m_d^2\right]^3}+\ldots\,,\\
\omega^{(\pm)}_{F,4}(\ell)=\omega^{(\pm)}_{F,5}(\ell)&=p_F(\ell)\pm m_d
+\frac{M^2}{2[2p_{F}(\ell)\pm m_d]}\pm\frac{[2p_{F}(\ell)\mp m_d]M^4}{8m_d [2p_{F}(\ell)\pm m_d]^3}+\ldots
\,,
\end{align}
Expanding similarly the bosonic fluctuations~\eqref{eq_app_5field_omegaB} for $\ell \geq 1$ (the expansion of $\omega_{B,3}$ is singular at $\ell=0$) and evaluating the sum~\eqref{eq_5field_Casimir} in dimensional regularization, \eqref{eq_5field_Delta0_small} follows.

To obtain the result~\eqref{eq_5field_Delta_large}, we isolated explicitly the divergent contribution in the sum. Noting that the first two modes in~\eqref{eq_app_5field_omegaB} and the two modes in~\eqref{eq_app_5field_omegaF_12} sum to a vanishing contribution in dimensional regularization, we recast~\eqref{eq_5field_Casimir} as
\begin{equation}\label{eq_app_5field_Casimir}
    \hat\Delta_0=\frac{R}{2} \sum_{k=3}^5 \sum_{\ell=1}^{\infty} \tilde{\sigma}(\ell)
    +\sigma_0
    \;,
\end{equation}
where $\tilde\sigma$ denotes the summand in $d=4$ from which we subtracted the terms of the $\ell\rightarrow\infty$ expansion whose sum does not converge in $d=4$; explicitly we find:
\begin{equation}
\tilde{\sigma}(\ell)=
\frac{R}{2}\sum_{k=3}^5\left[n_B(\ell) \omega_{B,k}(\ell)-\frac12 \sum_{ \pm} n_F(\ell) \omega^{( \pm)}_{F,k}(\ell)\right]+
\frac{9 \ell^2}{4}+\frac{21 \ell}{4}+3+\frac{R^2M^2}{4}\,,
\end{equation}
that can be checked to be $O\left(\ell^{-2}\right)$ for $\ell\rightarrow\infty$. In~\eqref{eq_app_5field_Casimir} $\sigma_0$ is the contribution of the (regularized) divergent sum and the $\ell=0$ states
\begin{equation}
\sigma_0=3+\frac{3 (RM)^2}{8}+
\frac{R}{2} \sum_{k=3}^5 \left[ \tilde{\omega}_{B,k}(0)-4N_f \sum_{ \pm}  \tilde{\omega}^{( \pm)}_{F,k}(0)\right]\,.
\end{equation}
To obtain the scaling dimension for large $g^2 Q$ we then separated the sum~\eqref{eq_5field_Casimir} into two contributions with a cutoff $a M R$ with $M\gg 1/R$ and $a\ll 1$ as in appendix $D$ of \cite{Cuomo:2021cnb}. The sum over the low $\ell$'s is evaluated expanding the dispersion relations at $R^2 M^2\gg \ell\sim O(1)$, while the sum over the large $\ell$'s is computed expanding for $\ell\sim R M\gg 1$ and using the Euler-Maclaurin formula. The details are essentially identical to~\cite{Cuomo:2021cnb} and we do not report them here.

\section{\texorpdfstring{$\mathbb{CP}^{N_f-1}$}{CPNf} Casimir energy}\label{app:CP1casimir}

\subsection{Special case: \texorpdfstring{$N_f=2$}{Nf=2}}

For the special case $N_f=2$, we have $\mathbb{CP}^1=S^2$. The bosonic part of the EFT is just 
\begin{equation}\label{eq:CP1_EFT}
    \mathcal{L}=\frac{1}{2}\left(m_d^2 \Phi^2 -\Phi\nabla^2 \Phi\right)+\frac{c}{2}\Phi^2 \left((\partial\varphi)^2 +\sin^2 \varphi(\partial\theta)^2 \right)\;.
\end{equation}
Focus on the $U(1)$ isometry which takes $\theta\to \theta+\alpha$. There is a solution to the EOMs of the form
\begin{equation}\label{eq:CP1_sol}
\Phi=v\,,\quad\varphi=\varphi_{0}=\pi/2\,,\quad\theta=i\mu t\,;
\end{equation}
this is an $S^2$ geodesic and so solves the EOMs for $\theta,\,\varphi$. The $\Phi$ EOM reduces to
\begin{equation}
\partial^2 \Phi=\left(m_d^2 +c\left(\partial\theta\right)^2 \right)\Phi\,,
\end{equation}
which fixes
\begin{equation}
\mu=\frac{m_d}{\sqrt{c}}\,.
\end{equation}

Now we can find the leading contribution to the energy by directly plugging this into the Hamiltonian:
\begin{equation}E=\frac{1}{2}\left(m_d^2 v^2 \right)+\frac{c}{2}v^2 \mu^2 =m_d^2 v^2\;.
\end{equation}
The charge is $q=v^2 \sqrt{c}m_d$
and so we find
\begin{equation}
E=\frac{m_d^2 q}{m_d\sqrt{c}}=\frac{m_d}{\sqrt{c}}q\;,
\end{equation}
which has the expected linear behavior.

Next we find the subleading correction to the energy from the one-loop Casimir energy. We expand around the saddle point using
\begin{equation}\label{eq:app_sol}
\Phi=v+\Sigma\,,\quad\theta=i\mu t+\Theta\,,\quad\varphi=\pi/2+\phi\,,
\end{equation}
and expand the action to second order in the fluctuations. The quadratic terms are
\begin{equation}
\mathcal{L}\supset-\frac{1}{2}\Sigma\nabla^2 \Sigma+\frac{cv^2 }{2}\left(\partial\Theta\right)^2 +2\sqrt{c}m_dvi\Sigma\partial_{0}\Theta+\frac{cv^2 }{2}\left(\partial\phi\right)^2 +\frac{v^2 m_d^2 }{2}\phi^2\;, \end{equation}
giving the mass matrix
\begin{equation}(\Sigma\;\;\Theta\;\;\phi)\left(\begin{array}{ccc}
\frac{1}{2}\left(J_{\ell}^2 -\omega^2 \right) & \sqrt{c}m_dv\omega\\
\sqrt{c}m_dv\omega & \frac{cv^2 }{2}\left(J_{\ell}^2 -\omega^2 \right)\\
 &  & \frac{1}{2}cv^2 \left(J_{\ell}^2 -\omega^2 \right)+\frac{v^2 m_d^2 }{2}
\end{array}\right)\begin{pmatrix}\Sigma\\ \Theta\\ \phi\end{pmatrix}\;.
\end{equation}
The dispersion relations are
\begin{equation}
\begin{split}
    R\omega_{\pm}	&=\frac{1}{2}\left(d-2\pm(d+2\ell-2)\right)=\begin{cases}
\ell+d-2\\
\ell\,,
\end{cases}\\
R\omega_{\phi}	&=\frac{\sqrt{c\ell(d+\ell-2)+\frac{1}{4}(d-2)^2 }}{\sqrt{c}}\,.
\end{split}
\end{equation}
The casimir energy is 
\begin{equation}\frac{1}{2}\sum_l n_\ell \omega_\ell\,,
\end{equation}
with $n_\ell=1+2\ell$ in 3d. This sum diverges so we compute it using dimensional regularization, following \cite{Badel:2019oxl}. The contributions from $\omega_{\pm}$ vanish, so we only need to take into account the contribution from $\omega_\Phi$. The result is displayed in figure \ref{fig:CP1}.

Next we show that the fermions do not contribute to the Casimir energy. The relevant terms are those described in \eqref{eq_EFT_fermion_3dSUSY} and take the explicit form (up to an overall factor)
\begin{equation}
i\left(D_{\alpha\beta}\psi^{i\beta}\psi^{j\alpha}-\psi^{i\alpha}D_{\alpha\beta}\psi^{j\beta}\right)\,,
\end{equation}
where
\begin{equation}
\psi^{i\alpha}D_{\alpha\beta}\psi^{j\beta}=\psi^{i\alpha}\left(g_{ij}\sigma^\mu\pd_\mu+\partial_{i}g_{kj}\sigma^\mu_{\alpha\beta}\pd_\mu\phi^{k}\right)\psi^{j\beta}\,,
\end{equation}
and $g$ is the metric for the NLSM above,
\begin{equation}
g_{ij}=\left(\begin{array}{ccc}
\frac{1}{2}\\
 & \frac{c\Phi^{2}}{2}\\
 &  & \frac{c\Phi^{2}}{2}\sin^{2}\varphi
\end{array}\right)\,.
\end{equation}
Evaluating this on the solution \eqref{eq:app_sol} we find the quadratic terms
\begin{equation}
\psi_\phi\partial\psi_\phi 
+cv^2\psi_\theta\partial\psi_\theta +cv\psi_\phi \partial\theta\psi_\theta+\psi_\varphi\partial\psi_\varphi\,,
\end{equation}
since $\psi_\varphi$ appears as a free massless fermion it does not contribute to the Casimir energy and we ignore it. The remaining dispersion relations are
\begin{equation}\label{eq:app_ferm_disp}
    \omega_F^{(\pm)}(\ell)=p_F(\ell)\pm \sqrt{c}\mu=p_F(\ell)\pm \frac{1}{2R}\,.
\end{equation}
Since $\omega_F^{(\pm)}(\ell)$ are always positive, the fermions do not contribute to the Casimir energy. The full result for the Casimir energy is then just the bosonic contribution computed above.

\subsection{General \texorpdfstring{$N_f$}{Nf}}\label{app:general_nf}

Now we perform the calculation for general $\mathbb{CP}^{N_f-1}$. First we must choose a $U(1)$ symmetry. Writing $z_{1}=re^{i\theta}$, we can choose the $U(1)$ symmetry which shifts $\theta$. We guess a solution to the EOMs of the form
\begin{equation}
    \Phi=v\,,\quad\theta=i\mu t\,,\quad r=r_{0}\;,
\end{equation}
and where all $z_M=0$ for $M>1$. This is a geodesic for $r_0=1$, and so solves the EOMs for the $\mathbb{C}^{N_f-1}$ coordinates. To fix $\mu$ we use the EOM for $\Phi$, which fixes 
\begin{equation}
\mu=\frac{m_d}{\sqrt{c}}\;.
\end{equation}

Now we look at small fluctuations:
\begin{equation}
    \Phi=v+\Sigma,\quad r=r_{0}+R\,,\quad\theta=i\mu t+\Theta\,,\quad\quad z_{M}=Z_{M}\,,
\end{equation}
where $M>1$. Expanding the action to quadratic order, we find that the the mass matrix for $\Sigma,R,\Theta$ is independent of the number of $Z_M$'s and so is the same as in the $CP^{1}$ case, and in particular their contribution to the Casimir energy is identical. We next have to add the contribution from the $Z_M$'s. They contribute as $N_f-2$ copies of the dispersion relations
\begin{equation}
    \omega_{Z,\pm}(\ell) = \frac{1}{2} \left( \sqrt{\mu^2 + 4\ell(\ell + d - 2)} \pm \mu \right)\;.
\end{equation}
When we sum over these modes, the contribution from $\pm \mu$ cancels, and we are left with the contribution of $2(N_f-2)$ real massive fields of mass $\frac12\mu=\frac{m_d}{2\sqrt c}$. The total Casimir energy is then the sum of the contribution from the $N_f=2$ sector above together with these $2(N_f-2)$ relativistic modes. Relativistic modes always contribute a non-positive value to the Casimir energy and so these can only lower the Casimir energy.

To finish the calculation we also need to find the fermionic contribution. Expanding the fermionic part of the EFT to quadratic order and ignoring free massless fermions, we find only one nontrivial dispersion relation, which is identical to \eqref{eq:app_ferm_disp}. As a result we again find that the fermions do not contribute to the Casimir energy.

\section{Correlation functions of a free real scalar field from semiclassics}\label{app_free_semiclassics}

In this section we show how to generalize the semiclassical approach of~\cite{Badel:2019oxl} to the calculation of correlation functions of heavy neutral operators in the theory of a massless real free field $\Phi$. This motivates considering the saddle-point~\eqref{eq_phi_cyl} in the dilaton EFT.

We consider the two-point function of the operator $\Phi^n$. This corresponds to the following path-integral
\begin{equation}
\langle\Phi^n(x_f)\Phi^n(x_i)\rangle=\int\mathcal{D}\Phi\,
\Phi^n(x_f)\Phi^n(x_i)e^{-\int d^dx\frac{1}{2}\left(\pd\Phi\right)^2}\,.
\end{equation}
Proceeding as in \cite{Badel:2019oxl}, we exponentiate the insertion and obtain the following modified action
\begin{equation}
S_{mod}=\int d^dx\frac{1}{2}(\pd\Phi)^2-n\log \Phi(x_f)-n\log\Phi(x_i)\,.
\end{equation}
We then consider the resulting saddle-point equation:
\begin{equation}
-\pd^2\Phi=\frac{n}{\Phi(x)}\left[\delta^d(x-x_f)+\delta^d(x-x_i)\right]\,.
\end{equation}
The solution to this equation formally reads
\begin{equation}\label{eq_app_free_flat_sol}
\Phi(x)=\alpha \,G(x-x_i)+\beta\, G(x-x_f)-(\alpha+\beta)G(0)\,,
\qquad G(x)=\frac{1}{(d-2)\Omega_{d-1}|x|^{d-2}}\,,
\end{equation}
where the coefficients $\alpha$ and $\beta$ satisfy
\begin{equation}
\alpha \beta=\frac{n}{G(x_{if})-G(0)}\,,
\end{equation}
and $x_{if}=x_i-x_f$. Note that the solution therefore admits a free parameter; the physical significance of the latter will become clear once we discuss the map to the cylinder. In~\eqref{eq_app_free_flat_sol} we implicitly assumed that we work within a regularization scheme in which $G(0)$ is finite. The necessity of specifying a regularization scheme is unsurprising, since also in the diagrammatic approach we need to specify a prescription for the tadpole diagrams arising from contractions of fields at the same point. In mass-independent schemes that preserve the conformal symmetry of the free theory, such as dimensional regularization, $G(0)\rightarrow 0$ by dimensional analysis. In the following we will assume that we work in such a scheme, for which the solution simplifies and reads
\begin{equation}\label{eq_sol_flat2}
\Phi(x)=\sqrt{\frac{n}{G(x_{if})}}\left[e^{\tau_0} G(x-x_i)+e^{-\tau_0} G(x-x_f)\right]\,,
\end{equation}
where $\tau_0$ is arbitrary.

It is now simple to verify that the saddle-point~\eqref{eq_sol_flat2} gives the expected result for the two-point function for $n\gg 1$. Indeed, plugging the solution in the classical action we obtain
\begin{equation}
\langle\Phi^n(x_f)\Phi^n(x_i)\rangle
\simeq e^{-S_{mod}\vert_{saddle}}=
e^{n\log n-n} [G(x_{if})]^n \approx n!\, [G(x_{if})]^n\,,
\end{equation}
where we used the Stirling formula in the last step. 

Let us finally map the solution~\eqref{eq_sol_flat2} to the cylinder, such that the corresponding Euclidean time stretches between the two operators insertions at $\tau=\pm\infty$ in terms of the Euclidean cylinder time coordinate. Introducing a subscript to distinguish between flat space and cylinder fields, the map is given by
\begin{equation}
\Phi_{flat}(x)=\left(\frac{|x_{if}|R}{|x-x_i||x-x_f|}\right)^{\frac{d-2}{2}}\Phi_{cyl}(\tau,\hat{n})\,,\qquad
e^{\tau/R}=\frac{|x-x_i|}{|x-x_f|}\,,
\end{equation}
where $R$ is the cylinder radius and $\hat{n}$ specifies the position on the sphere. We therefore obtain the Euclidean profile
\begin{equation}
\Phi_{cyl}(\tau,\hat{n})=\sqrt{\frac{n}{(d-2)\Omega_{d-1}R^{d-2}}}\left[e^{-m_d\tau+\tau_0}+e^{m_d\tau-\tau_0}\right]\,,
\end{equation}
where $m_d=\frac{d-2}{2R}$. Wick rotating $\tau\rightarrow i t$ and $\tau_0\rightarrow i t_0$, and removing the subscript, we obtain the profile~\eqref{eq_phi_cyl} in the main text. Note that the arbitrary parameter of the solution now admits a simple interpretation as the zero-mode parametrizing the origin of the oscillation period of the solution.

\bibliography{refs}
	\bibliographystyle{JHEP.bst}

\end{document}